\documentclass[final,3p,twocolumn,times]{elsarticle}
\usepackage[utf8]{inputenc}
\usepackage{ulem}
\usepackage{tabularx}
\usepackage{amsmath}
\usepackage{graphicx}
\usepackage{subfig}
\usepackage[hyphens]{url}
\usepackage{float}
\usepackage{comment}
\usepackage{flushend}
\usepackage[inline]{enumitem}

\journal{Computer Networks}

\begin{document}


\begin{frontmatter}
\title{Impact of the COVID-19 pandemic on the Internet latency: a large-scale study}

\author[dip]{Massimo~Candela}
\ead{massimo.candela@ing.unipi.it}
\author[iit]{Valerio~Luconi}
\ead{valerio.luconi@iit.cnr.it}
\author[dip]{Alessio~Vecchio\corref{cor1}}
\ead{alessio.vecchio@unipi.it}

\cortext[cor1]{Corresponding author}
\address[dip]{Dip. di Ing. dell'Informazione\\Universit\`a di Pisa\\Largo L. Lazzarino 1, 56122 Pisa, Italy}
\address[iit]{Istituto di Informatica e Telematica\\Consiglio Nazionale delle Ricerche\\Via G. Moruzzi, 1, 56124 Pisa, Italy}

\date{May 2020}

\begin{abstract}

The COVID-19 pandemic dramatically changed the way of living of billions of people in a very short time frame. In this paper, we evaluate the impact on the Internet latency caused by the increased amount of human activities that are carried out on-line. The study focuses on Italy, which experienced significant restrictions imposed by local authorities, but  results about Spain, France, Germany, Sweden, and the whole of Europe are also included. The analysis of a large set of measurements shows that the impact on the network can be significant, especially in terms of increased variability of latency. In Italy we observed that the standard deviation of the average additional delay -- the additional time with respect to the minimum delay of the paths in the region -- during lockdown is $\sim3-4$ times as much as the value before the pandemic. Similarly, in Italy, packet loss is $\sim2-3$ times as much as before the pandemic. The impact is not negligible also for the other countries and for the whole of Europe, but with different levels and distinct patterns. 

\end{abstract}

\end{frontmatter}
\let\thefootnote\relax\footnotetext{This manuscript version is made available under CC-BY-NC-ND~4.0 license http://creativecommons.org/licenses/by-nc-nd/4.0/

First appearance on arXiv: 13 May 2020

Please, refer to the published version in Computer Networks, Elsevier. DOI: https://doi.org/10.1016/j.comnet.2020.107495}

\section{Introduction}

At the time of writing, the coronavirus disease (COVID-19) pandemic is still ongoing and billions of people are under some form of lockdown. The restrictions faced by citizens are more or less stringent, depending on the resolutions adopted by the different governments, but in many cases non-essential activities have been shut down and a large fraction of people is confined at their homes. Many activities that are normally carried out in physical presence are now taking place on-line. As a consequence, the amount of traffic on the Internet increased significantly during the last months. 

In this paper, we analyze the impact of the COVID-19 pandemic on the latency of the Internet. Latency is one of the major properties of the network and it is becoming every day more important, as several Internet applications are particularly sensitive to its fluctuations. Examples include on-line videogames~\cite{10.1145/1167838.1167860, 7008964}, video calls, VOIP~\cite{Janssen02:assessing}, and IP geolocation~\cite{CIAVARRINI201870, Ciavarrini17:smartphone,8685105}. We analyzed a large set of measurements, collected by means of the RIPE Atlas platform~\cite{staff2015ripe}, to better understand the effects on the network caused by this major change in the way we live. 

The analysis focuses on Italy which, in April 2020, has been under lockdown for more than a month, experiencing some of the strictest limitations enforced by authorities: all schools, universities, and non-essential shops are physically closed, and people are authorized to leave their homes only for undeferrable necessities. Distance learning and remote working were applied whenever possible, with a significant increase in usage of virtual-meeting and video-conference applications \cite{teams,zoom}.
Table~\ref{tab:events} summarizes the most important events which could have had an impact on the Italian Internet latency. As can be noticed, limitations to citizens have been introduced progressively. For this reason, the changes caused by the Italian lockdown are studied, in the remaining of this paper, by comparing the situation of the network in the  February 11-17 week with the March 10-16 week. The former represents the ``normal'' status of the network, as it is antecedent to all restrictive measures; the latter, on the contrary, comes just after the most restrictive limitations. The period in between corresponds to a transitory phase, where partial lockdowns start to impact the network performance. Hereafter, we will use W1 to indicate the baseline week, and W2 for the week just after the major lockdown event.

Besides Italy, we include a brief analysis also concerning Spain, France, Germany, Sweden, and the whole of Europe. Spain, France, and Germany have been characterized by restrictions similar to the Italian ones. Sweden instead decided not to impose a mandatory national lockdown. For Spain, France, and Germany W2 is shifted according to their major lockdown event (shown in Table~\ref{tab:events}). For the entire Europe the situation is more heterogeneous, as some countries were less hit by COVID-19 and thus adopted milder restrictions. For Sweden and Europe, W2 corresponds to the March 20-26 week, the last of our observation period which goes, overall, from February 11 through March 26. Results show that the impact is not always the same across the considered countries and on a European scale.

\begin{table*}[t!]
    \centering
    \caption{Events that could have had an impact on the latency in the European Internet (year 2020).}
    \begin{tabularx}{\textwidth}{|l|X|}
    \hline
    {\it Date}     &  {\it Event} \\
    \hline
    February 22 & Some Italian municipalities involved in the COVID-19 outbreak are quarantined ($\sim 50$ thousand people)\\
    \hline
    March 1 & A larger number of Italian municipalities are quarantined (the so called ``red zone''), while in others (the ``yellow zone'') schools, cinemas, sport events, and other gatherings are suspended\\
    \hline
    March 4 & Schools and universities are shutdown nationwide\\
    \hline
    March 7  & Several northern Italian regions are put on lockdown (approximately 16 million people)\\
    \hline
    March 9 & All Italy is put on lockdown\\
    \hline
    March 14 & Spain is put on lockdown\\
    \hline
    March 16 & France is put on lockdown\\
    \hline
    March 22 & Germany is put on lockdown\\
    \hline
    \end{tabularx}
    \label{tab:events}
\end{table*}

The contribution of this paper can be summarized as follows:

\begin{itemize}
    \item Some statistics have been recently released by Internet Service Providers (ISPs), and other players of the Internet ecosystem, about the increased amount of traffic they have been exposed to because of the COVID-19 pandemic. However, a picture that leaves aside the very specific points of view of the single operators is still missing. This study provides a more global view, not polarized by the single operator's perspective. In addition, most of the statistics released by operators concern the amount of traffic, with limited (or absence of) information about latency.
    
    \item The amount of measurements analyzed is large, thus providing solid foundations for the included statistics. Moreover, besides the sheer number, we decompose the impact on delay according to the most relevant factors, including the time of the day, the type of target (belonging to a content delivery network or not), and the version of the Internet Protocol. In addition to ICMP-based delay, we provide information also on packet loss and path changes, as they are, more or less, related with Internet latency, and on HTTP-based latency.
    
    \item Besides Italy, results also concerning Spain, France, Germany, Sweden, and the whole of Europe are included. Since measurements have been collected using a single platform, results obtained for the different countries can be compared without the possible bias introduced by the adoption of multiple and heterogeneous systems. 
    
\end{itemize}

Results show that, in Italy, lockdown impacted the latency of the network, especially in terms of increased variability. Effects are more evident during the evening, suggesting that latency is more negatively affected by the increased traffic due to recreational activities rather than remote working or distance learning. For the other European countries included in the study, the impact is milder in Germany and France, and similar to the Italian one in Spain and Sweden.

The remaining of the paper is organized as follows: in Section~\ref{sec:relwork}, we summarize the most significant work concerning the detection and analysis of large anomalies occurred in the Internet; Section~\ref{sec:data} describes the data collection phase; in Section~\ref{sec:method}, the method we followed to compute the performance indexes is explained; the main characteristics of the datasets are illustrated in Section~\ref{sec:preliminary}, together with a preliminary analysis; Section~\ref{sec:result} contains the results on the Italian Internet latency from different perspectives (type of measurements, hour of the day, IPv4 vs IPv6, etc), whereas Section~\ref{sec:europe} shows the results concerning the above-mentioned countries and the whole of Europe (with less details compared to Italy); Section~\ref{sec:conclusion} concludes the paper.

\section{Related work}
\label{sec:relwork}

There is an extensive body of literature about anomalies on the Internet. However, the focus is, usually, on the design of techniques aimed at detecting the occurrence of an anomaly (for instance~\cite{10.5555/1251086.1251118, 6932930, Chen16:measurement,XIA201815}). In this paper, we do not try to define another detection technique, as the impact of the COVID-19 pandemic on the network is evident. We strive to provide a view of the effects of the pandemic at a large-scale, including a quantitative evaluation of its main characteristics as seen from the perspective of latency.

When natural events assume catastrophic proportions, their effects might be observed also on the performance and reliability of the Internet. During the last 20 years, several studies have focused on the impact on the Internet of a few catastrophic natural events, such as earthquakes and hurricanes. 
In~\cite{Kitamura07:experience}, authors studied the effects of the Taiwan 2006 earthquake on the Asian Internet, from the viewpoint of interdomain routing and traffic in research and educational networks. The earthquake damaged submarine fiber cables causing the failure of multiple links and unavailability of several routes. The paper shows that through automatic rerouting performed by the BGP protocol and traffic engineering via backup or redundant paths, the connectivity could be restored within a matter of hours, even if with some loss of performance. Other studies reported more significant damage and instability for longer periods~\cite{Popescu07:quaking}. Cho et al. conducted a similar study on the Japan 2011 earthquake and subsequent tsunami in~\cite{Cho11:japan}. The study analyzes the impact on the Internet as seen by a local ISP, analyzing both traffic and routing. Due to link failures, the Internet experienced traffic drops and subsequent peaks mainly due to the reconfiguration of content distribution networks. However, thanks to planned backup and redundancy, the Internet proved to be resilient to such an event, and was affected only locally and with minimal damage. Another study analyzed the latency variations due to the same event~\cite{Japanearthquake}, as seen from PingER monitors~\cite{Matthews00:pinger}. The study shows that, after the earthquake, latencies from some monitors experienced a significant increase for a limited amount of time. These studies point out the importance of maintaining redundant Internet routes, even if at a cost, as they can be extremely useful when disruptive events occur.

Similar work was conducted on the effects caused by severe weather conditions, such as the 2012 hurricane Sandy. In~\cite{Heidemann12:preliminary}, authors used ping to test the reachability of edge networks, to discover that the number of outages in the areas affected by the hurricane Sandy significantly increased during and after the hurricane. From the interdomain routing point of view, Aben shows that a significant portion of traffic was rerouted around the affected area, again demonstrating the Internet resilience and the importance of a redundant structure~\cite{Aben12:hurricane}. Similar outages can be observed also at a smaller scale, as pointed out in~\cite{Schulman11:pingin}. The study uses ping to show that reachability issues can occur at residential hosts in case of moderately bad weather conditions (e.g. thunderstorms). However, unlike large-scale outages, on the small scale the degree of redundancy is generally not sufficient to cope with these events.

The vast majority of the studies concerning the impact of natural events on the Internet found in scientific literature tackle the problem from a reachability, routing, or traffic point of view. Very few analyzed the experienced latency increase in such events as we do for the COVID-19 pandemic. However, we believe that studying latency is of paramount importance, as it can give an indication of the perceived quality of service by end users. In addition, none of the previously studied events reached the size of the COVID-19 pandemic, in both space and time. At the time of writing, the COVID-19 pandemic has been impacting the lives of billions of people all around the globe for several months, while in the other analyzed natural events the impact of the Internet was limited in time or circumscribed to a relatively small geographic area. 

Since the COVID-19 pandemic has started only few months ago, there is still little scientific literature focusing on its impact on the performance of the Internet. However, several network operators, content providers, and Internet eXchange Points (IXPs), released reports about the increased usage of their infrastructure, which we review in the following. Cloudfare reported statistics about the traffic increase towards their servers placed in Seattle, Northern Italy and South Korea~\cite{cloudflare}. In particular, Cloudfare reports an increase of 30\% of the overall traffic in Northern Italy and a reduction of the traffic coming from fitness trackers perhaps reflecting the scarce mobility induced by social distancing. Similarly, Fastly reported traffic and download speeds towards their servers~\cite{fastly}. Various countries were considered, including Italy, which had a 109.3\% increase in terms of traffic and a 35.4\% decrease in terms of download speed on average. DE-CIX, one of the world's biggest IXPs, reported a new traffic world record of 9.1 Tbps~\cite{decix}, as well as a 50\% increase in video conferencing traffic and 25\% of social media traffic. Finally, the Organisation for Economic Co-operation and Development (OECD) released a report that aggregates the various traffic increase information reported by Internet operators in a single document~\cite{oecd}. Such report highlights some important numbers, among which: (i) an increase up to 60\% of traffic reported in many IXPs and ISPs; (ii) an increase up to 24 times higher of the volumes of traffic for video conferencing platforms.

All these reports focus on the increase of Internet traffic and not on latency. We believe that latency is extremely important as, without information about capacities, traffic itself cannot be used for an estimation of the perceived performance of the Internet by end users. Moreover, as already mentioned, all the currently available studies are limited to the boundaries of the organizations that provided them.

One of the few papers focusing on this topic, from a scientific perspective, is the one by Favale et al.~\cite{FAVALE2020107290}, where the impact of the COVID-19 pandemic is observed from the campus network of an Italian University. Favale et al. highlight a 10 time decrease in incoming traffic and an increase of 2.5 times in outgoing traffic, as a consequence of remote learning activities. Moreover, using passive measurements collected using Tstat~\cite{7876976} and application logs, they studied the fruition and performance of their in-house distance learning system.

\section{Collection of data}
\label{sec:data}

The raw latency data used for this study was collected by RIPE Atlas~\cite{staff2015ripe}. We then filtered and enriched such data as detailed in the following subsections. 
RIPE Atlas is a globally distributed Internet measurement platform that produces more than $10\,000$ measurements per second \cite{atlas}. Among the open platforms aimed at measuring the Internet, RIPE Atlas is the one with the largest number of vantage points \cite{bajpai2015survey}, and it has a massive presence in Europe.  

RIPE Atlas automatically carries out \textit{Anchoring Measurements} (AMs), where the set of targets is pre-defined. In particular, a large set of devices called \textit{probes} periodically perform measurements towards other devices called \textit{anchors}. Anchors are usually hosted in IXPs, in the operation centers of ISPs, or in data centers. Hence, they enable monitoring of the core infrastructure of the Internet.
The results produced by AMs have been extensively used for both research (see, for instance, \cite{bajpai2017vantage}, \cite{fontugne2017pinpointing}, \cite{shah2017disco}, \cite{vaton2017joint}, \cite{shah2018systems}) and operational purposes (for example DNSMON \cite{dnsmon}, a service aimed at monitoring the worldwide core DNS infrastructure).
Additionally, RIPE Atlas allows its users to collect measurements towards arbitrary targets. Results of \textit{User-Defined Measurements} (UDMs) are stored in a database from where they are accessible to the public (access is not restricted to the experimenters who triggered them).

RIPE Atlas, for its measurements, relies on classical network tools. The latency from an Atlas node to a target is estimated by means of the ping tool which, as known, makes use of ICMP echo requests and echo replies. For AMs, ping is launched to collect three Round Trip Time (RTT) values. For UDMs, the default number of collected RTT values is again three, but this number can also be changed by the experimenter. 

Our analysis of the impact of COVID-19 pandemic relies on the results generated by both AMs and UDMs.

\subsection{AM-derived dataset}
Starting from the dates of the events reported in Table~\ref{tab:events}, we considered all AMs comprised in the interval from the 11th of February to the 26th of March 2020. The set of probes and anchors involved in the measurements is stable, with little variations caused by the possible temporary unavailability of probes. Since AMs are performed periodically and for the entire period of study, they provide information on the Internet latency from a stable point of view. However, to be sure to eliminate measurements occurring in a short time frame and thus not covering appropriately the observation period, we further filtered the selection to contain only the measurements concerning source-target pairs that produced successful results in at least 40 different days (90\% of the total time frame).

The position of source and target nodes is fundamental to analyze the impact on a country-level basis. For AMs, the position of both source and target nodes is well-known, as such information is provided for each node participating in the platform. 
We use such information to select a subset of the ping measurements having both source and target in Europe. 

From now on we will refer to such a subset as the AM-derived dataset (AMD).
AMD is composed of more than $11.7$ billion RTT values generated by $303\,603$ source-target pairs during the monitored time period.

\subsection{UDM-derived dataset}

Users of the RIPE Atlas platform can define their own latency measurements according to their needs and interests. They can select the set of targets to be probed, define the periodicity of probing, and set the time-span of their measurement activities. 
In UDMs, targets frequently include the servers of major Internet companies, DNS servers, or privately owned network resources. As a consequence, the set of targets involved in UDMs is heterogeneous. Sources are always a subset of RIPE Atlas nodes, but the subset can be different from user to user.

For our analysis, we are interested in a subset of UDMs where measurement activities were possibly scheduled \textit{before} the COVID-19 outbreak in Europe and repeated periodically throughout the observation period.
To obtain such a subset, we adopted the following strategies. First, we extracted only periodic latency measurements, i.e. configured to be automatically repeated after a certain amount of time. 
Second, we discarded the measurements configured to collect less than three latency values per ping execution. 
Third, similarly to AMD, we filtered the selection to contain only the measurements concerning source-target pairs that produced successful results in at least 40 different days. 
Fourth, we restricted measurements to the ones targeting IP addresses in Europe. It is important to notice that in UDMs only the location of the source nodes is well-known, as it is provided by RIPE Atlas. Hence, for this step, we estimated the position of the targets by using RIPE IPmap~\cite{ripe-ipmap}. RIPE IPmap uses active geolocation, and it has been reported to be $100$\% accurate at continent level, $99.58$\% at country level~\cite{iordanou2018tracing}, and $80.3$\% at city level~\cite{ccripmap}. Additionally, we used MaxMind GeoLite2~\cite{maxmind} as a fallback tool, in case of failed IP geolocation with RIPE IPmap. Measurements where the target was not successfully geolocated using these two tools were discarded.
Finally, in some cases, we had to limit the amount of extracted information due to the almost unmanageable volume of data in the repository. In particular, when the number of targets in the geographic area of interest was too large, we randomly selected $10\,000$ targets and the analysis was restricted to them.
Even when we had to limit the number of targets, the number of sources from which measurements were started was not subject to any limitation. 

From now on we will refer to the dataset built according to the above-described procedure as the UDM-derived dataset (UDMD). UDMD is composed of more than $1$ billion RTT values generated by $671\,292$ source-target pairs during the observation time period.

\section{Method}
\label{sec:method}

Let $d$ be the delay of a given Internet path. It can be roughly expressed as $d = d_{tra} + d_{pro} + d_{que}$, where $d_{tra}$ is the transmission delay, $d_{pro}$ is the propagation delay, and $d_{que}$ is the time spent because of queues and processing at intermediate routers and target host. In a wide-area scenario like the one considered, $d_{pro}$ amounts to a significant fraction of the overall delay, as signals travel at approximately $200$ km/ms in fiber. Moreover, the only component that is going to be affected by increased traffic is $d_{que}$.
To isolate  $d_{que}$ from the other terms, the $d_{tra} + d_{pro}$ component can be estimated as the minimum delay observed in a set of latency measurements collected on a given path. The larger the number of collected samples, the better the quality of the estimate: intuitively, the collection of more samples increases the probability of finding lightly loaded network conditions.
This approach is not novel and it has been followed in other studies to characterize latency variations on a large scale. In \cite{10.1007/978-3-319-54328-4_13}, for each host-pair the difference between the maximum and minimum RTT observed in a time bin was calculated. Then, the evolution of the obtained difference values was used to investigate on transient congestion. The use of the minimum observed RTT as an approximation of the fixed delay associated with a path was adopted also in \cite{10.1145/948205.948241}, to study the variability in TCP connections.

Delay variation metrics are discussed in RFC 5481~\cite{rfc5481}, where one of the most widely implemented formulations is based on the use of the packet with the minimum delay in the sample as the reference packet. In particular, Packet Delay Variation is defined in RFC 5481 as the one-way delay of the considered packet minus the one-way delay of the packet with the lowest value for delay over the current test interval. Using the minimum delay as the basis for delay variation is the preferred method also according to the ITU-T Y.1540 Recommendation~\cite{itu1540} (albeit when considering 2-point packet delays, i.e. one-way delays computed by means of the absolute arrival time at destination minus the departure time at the source host). 

Starting from the above considerations, we defined the following method. 
Each source node, say $a$, measures the RTT towards a target, say $b$, using the ping command at time $t$, which produces a list of delay values $D_{t}^{ab} = \{RTT_{1}, RTT_{2}, ..., RTT_{n} \}$. Let us also define 
\begin{align*}
d_{min,t}^{ab} &= \min D_{t}^{ab}\\
d_{avg,t}^{ab} &= \text{avg}\: D_{t}^{ab}\\
d_{max,t}^{ab} &= \max D_{t}^{ab}    
\end{align*}

\noindent as the minimum, average, and maximum value found in the execution of the ping command at time $t$ from $a$ to $b$. First, we found the global minimum value observed for each source-target pair as 
\begin{equation*}
m^{ab} = min \bigcup\limits_{t \in O} D_{t}^{ab}
\end{equation*}

\noindent where $O$ is the entire period of observation.
Since the observation period we considered is relatively long, it is possible that some path changes occurred, i.e. that the set of traversed routers was not always the same. This is not a problem, provided that some of the above concepts are reformulated appropriately. In a scenario where path changes occur, $m^{ab}$ is an estimate of the transmission and propagation delay of the \textit{best} path between $a$ and $b$, among all the paths followed during the observation period.

Then, we used $m$ as a baseline to estimate the additional time experienced for every single pair of nodes. In particular, we computed 
\begin{align*}
q_{min, t}^{ab} &= d_{min,t}^{ab} - m^{ab}\\
q_{avg, t}^{ab} &= d_{avg,t}^{ab} - m^{ab}\\
q_{max, t}^{ab} &= d_{max,t}^{ab} - m^{ab}   
\end{align*}

\noindent for each couple $ab$ of nodes, and for each $t \in O$. The values of $q_{min, t}$, $q_{avg, t}$, and $q_{max, t}$ of all source-target pairs in the region of interest were then grouped in buckets with a duration of 30 minutes and averaged. This last step originates from our interest in evaluating the impact of the changed style of life on a large scale, for instance at the country level.
More formally, let us call $r_{min, k}$,  $r_{avg, k}$, and $r_{max, k}$ the average values obtained in the $k$th bucket $T_{k}$:
\begin{align*}
    r_{min, k} &= \text{avg}\:q_{min,t}^{ab}\\
    r_{avg, k} &= \text{avg}\:q_{avg,t}^{ab}\\
    r_{max, k} &= \text{avg}\:q_{max,t}^{ab}
\end{align*}

\noindent with $t \in T_k$ and across all $ab$ pairs located in the region of interest. In practice, $r$ represents the average additional time with respect to the best path ever followed during the whole observation period. Note that the average is computed for all measurements occurred in that specific bucket and across all source-target pairs located in the region of interest (e.g. all source-target pairs in Italy); as a consequence, the value of $r$ is generally greater than zero. To have $r_{min, k}$ equal to zero, all considered $ab$ couples should experience, almost simultaneously, their best RTT ($m^{ab}$) during the $k$th bucket, which is unlikely. 
The smallest observed delay found for a source-target pair ($m$) provides an indication of the ``uncompressible'' component of the delay for such pair and, as such, can be excluded in a study aimed at evaluating the impact of the pandemic.

\section{Characterization of the dataset and preliminary data analysis}
\label{sec:preliminary}

\begin{figure*}[t]
     \centering
     \subfloat[][Distances of source-target pairs in Italy]{
         \label{fig:cdf-distances-itit}
         \includegraphics[width=0.23\textwidth]{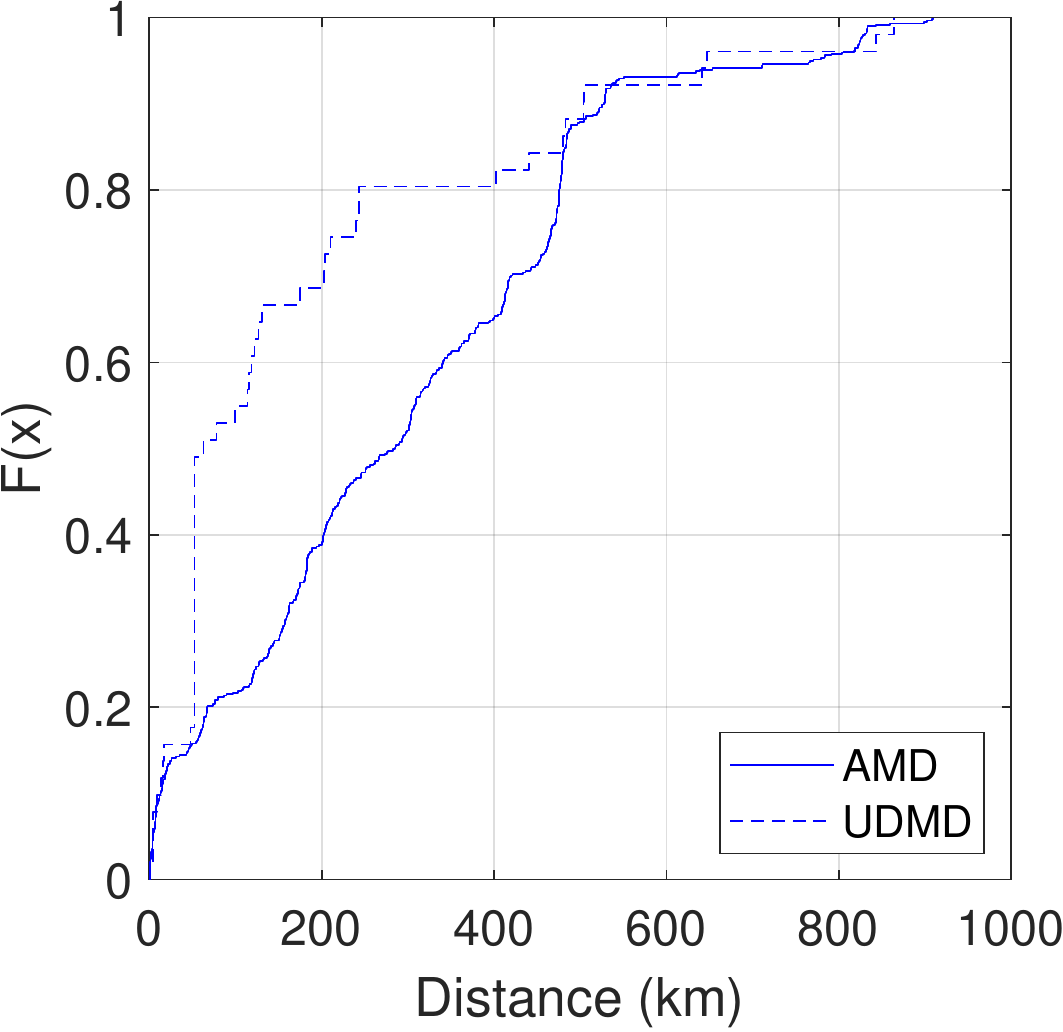}    
    }
    \hfill
     \subfloat[][Distances of source-target pairs in Europe]{
         \label{fig:cdf-distances-eueu}
        \includegraphics[width=0.23\textwidth]{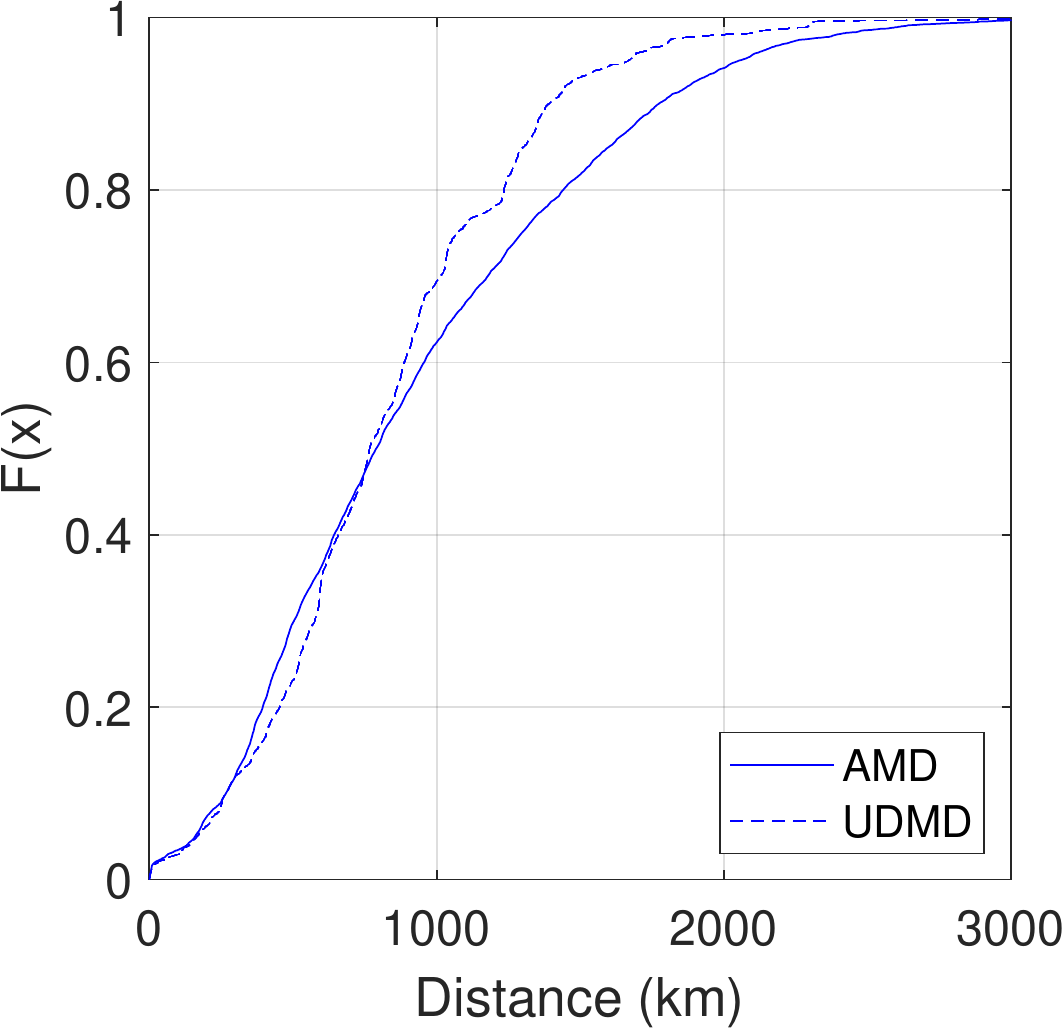}
     }
     \hfill
     \subfloat[][$m$ for source-target pairs in Italy]{
         \label{fig:cdf-m-itit}
         \includegraphics[width=0.23\textwidth]{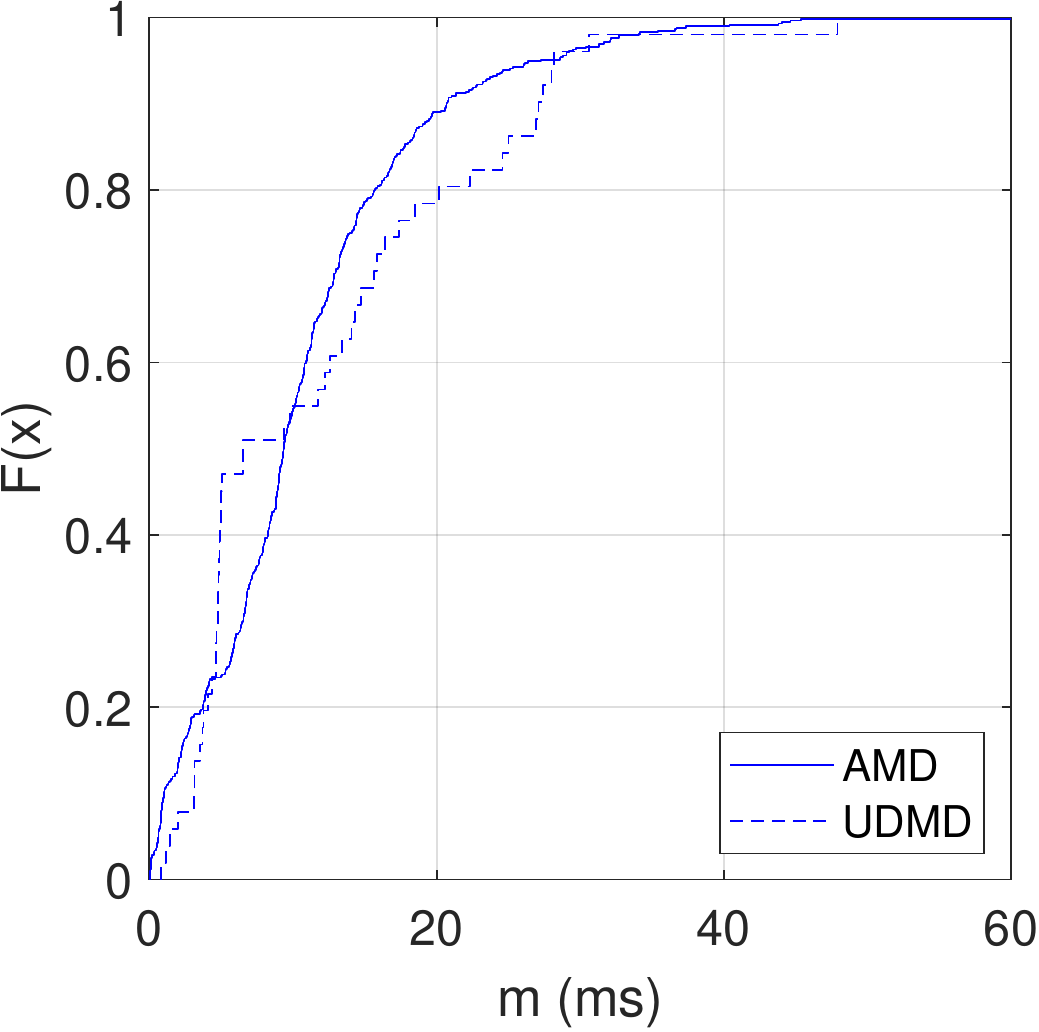}    
    }
    \hfill
     \subfloat[][$m$ for source-target pairs in Europe]{
         \label{fig:cdf-m-eueu}
        \includegraphics[width=0.23\textwidth]{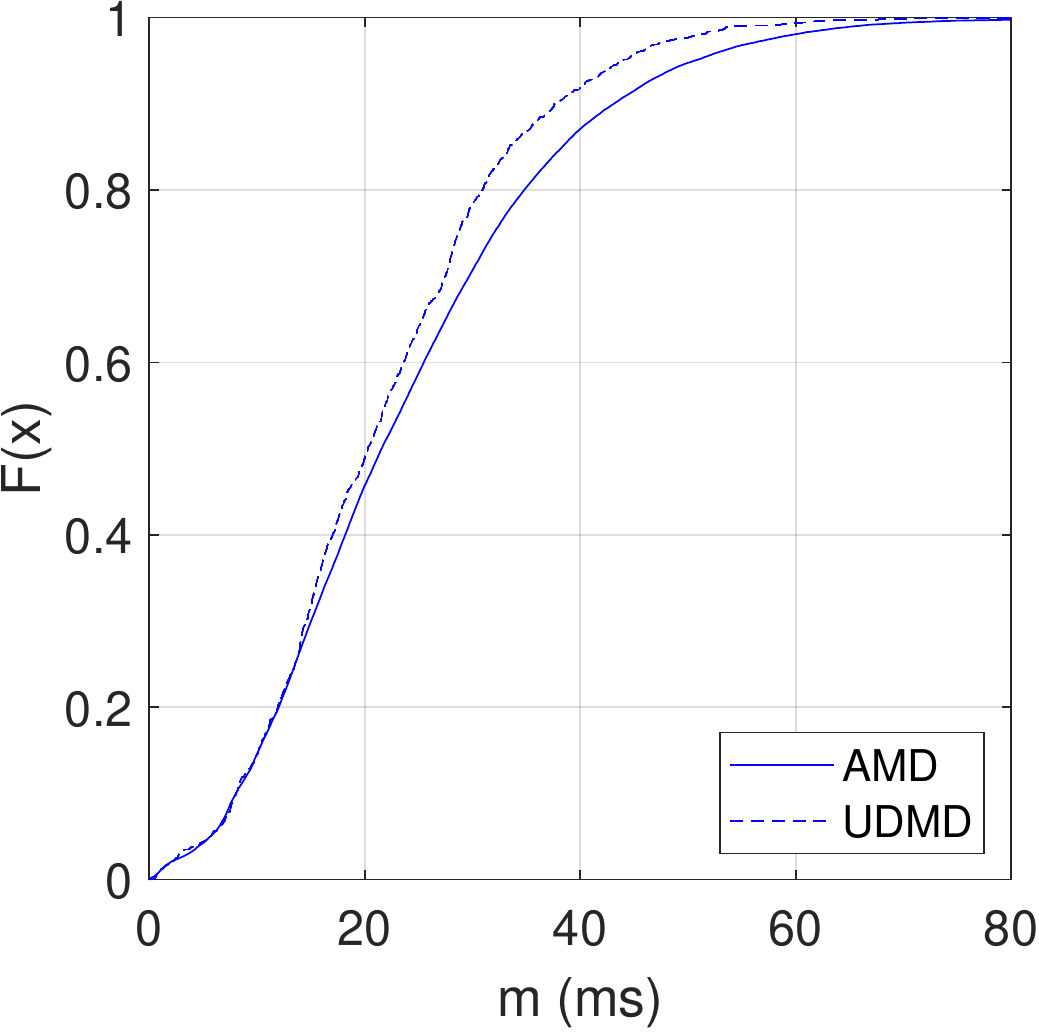}
     }
     \hfill
     \caption{Main characteristics of the Italian and European paths.\label{fig:allcdfs}}
\end{figure*}

In this section, we provide a picture of the datasets in terms of geographical extension and overall RTT dynamics.
Figure~\ref{fig:cdf-distances-itit} shows the distribution of distances of source-target pairs in Italy, for both AMD and UDMD. In detail, the distances between hosts are computed as Great Circle Distances (GCDs), i.e. the length of the shortest path on the surface of the Earth. Actual distances travelled by packets are generally larger than the corresponding GCDs because of circuitousness of Internet routes (deviations from shortest physical paths can be significant~\cite{matray2012spatial}). Distances span a couple orders of magnitude, from few kilometers, when source and target are in the same city or metropolitan region, to almost one thousand kilometers. Figure~\ref{fig:cdf-distances-eueu} shows the distribution of distances of European source-target pairs limited to 500 targets (but using all sources). For the European data, we had to limit the number of source-target pairs, because to compute the distance both source and target have to be located at city level, and active geolocation is expensive. For both Italian and European data, the distances reported in Figures~\ref{fig:cdf-distances-itit} and ~\ref{fig:cdf-distances-eueu} are the ones for which we have been able to compute the position of the involved nodes at city level (approximately 83-98\% of the considered couples). We believe that the distributions shown in Figure~\ref{fig:allcdfs} are reasonably close to the ones of the complete datasets. 

The distributions of minimum RTT value observed for each source-target pair (i.e. $m$) located in Italy and in the whole of Europe are shown in Figure~\ref{fig:cdf-m-itit} and ~\ref{fig:cdf-m-eueu}, respectively. The wide range of values, from few milliseconds up to $\sim40$ ms in Italy, and up to $\sim80$ ms in Europe, originates from the geographical extension of the considered areas. 

\begin{figure}[t!]
\centering
   \subfloat[][Raw RTT values and moving average (5 min)]{
         \label{fig:all_data_amd}
         \includegraphics[width=0.9\columnwidth]{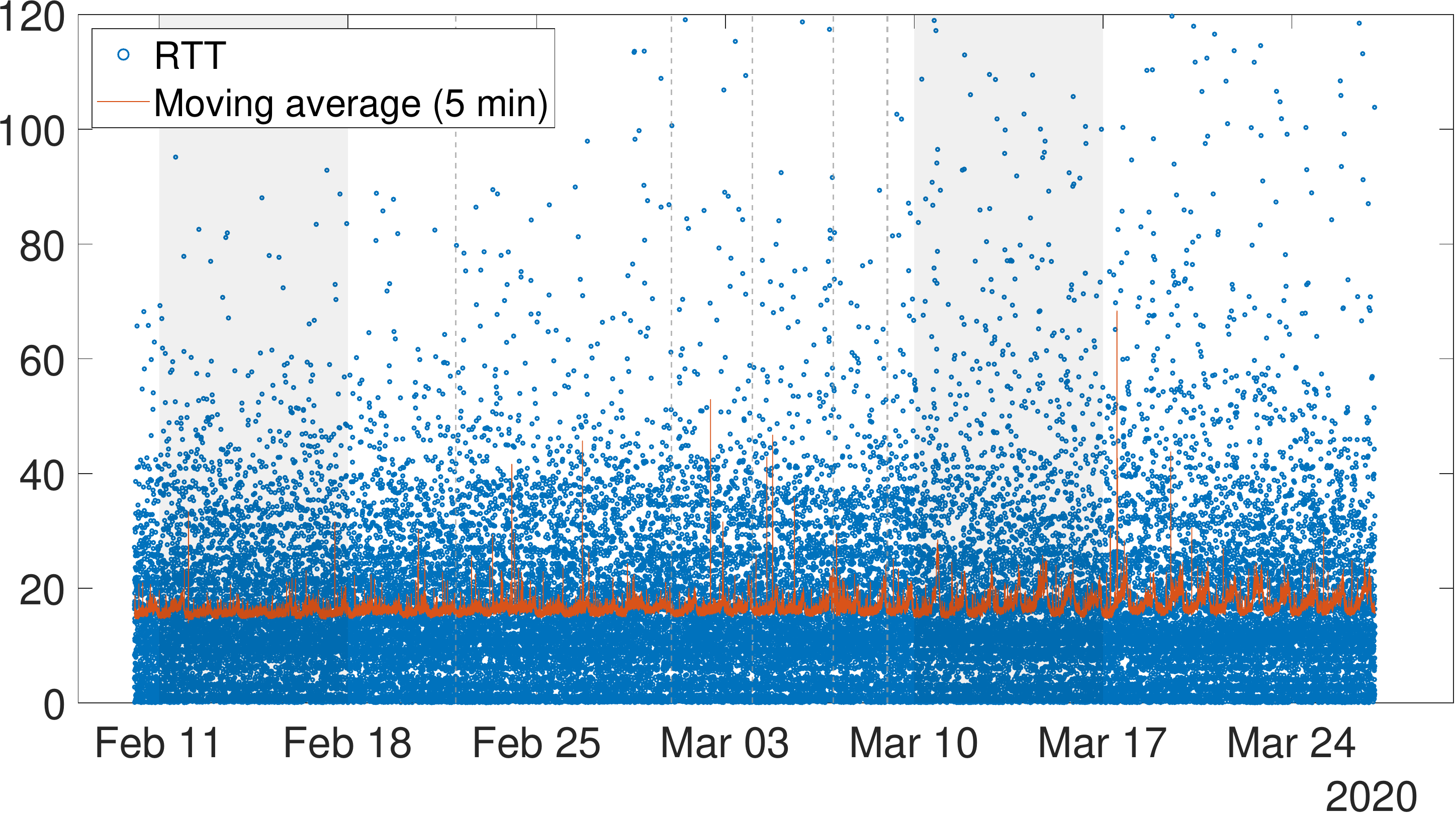}    
    }\\
    \subfloat[][$s_{min}$]{
         \label{fig:absolute}
         \includegraphics[width=0.9\columnwidth]{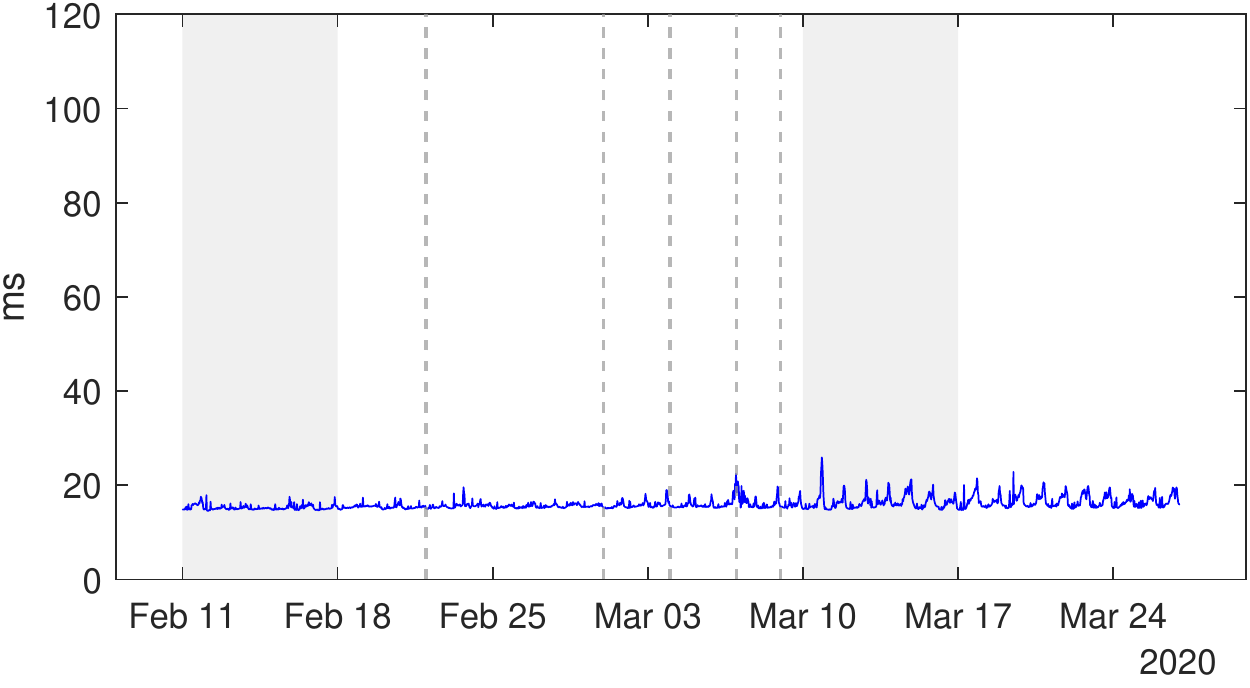}
     }\\
     \subfloat[][$s_{min}$ and $r_{min}$]{
         \label{fig:absolute_and_relative}
         \includegraphics[width=0.9\columnwidth]{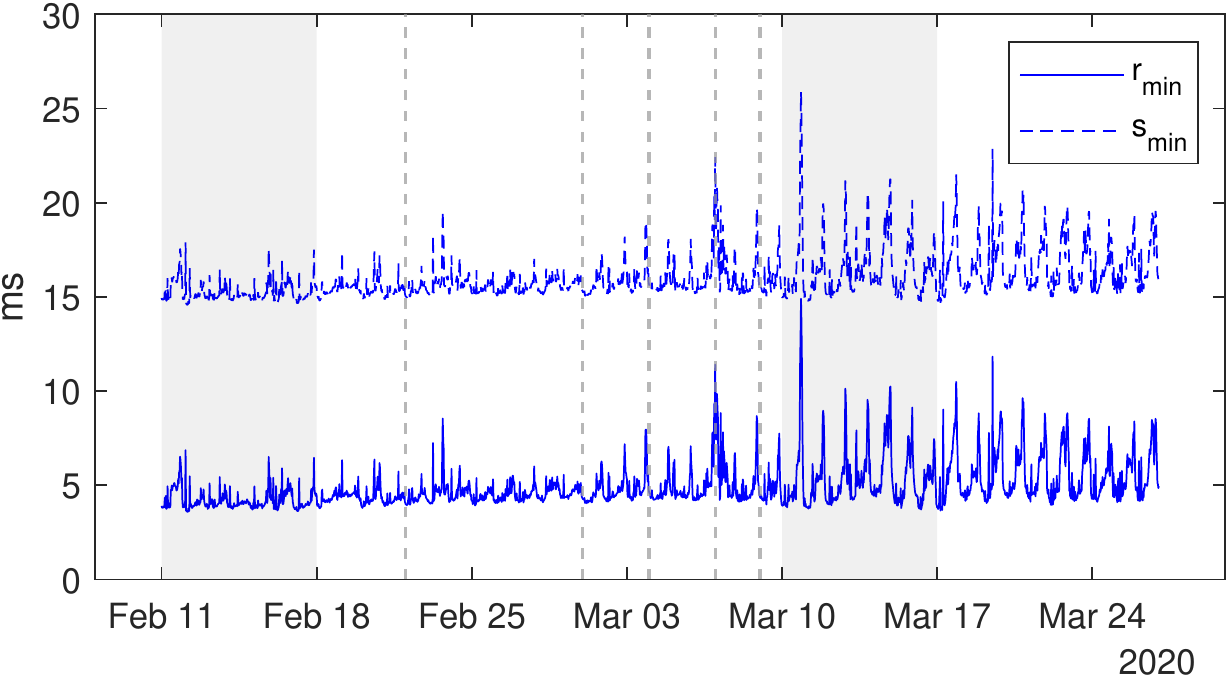}
     }\\
     \subfloat[][$q_{min}$, 95th percentile]{
         \label{fig:p95}
         \includegraphics[width=0.9\columnwidth]{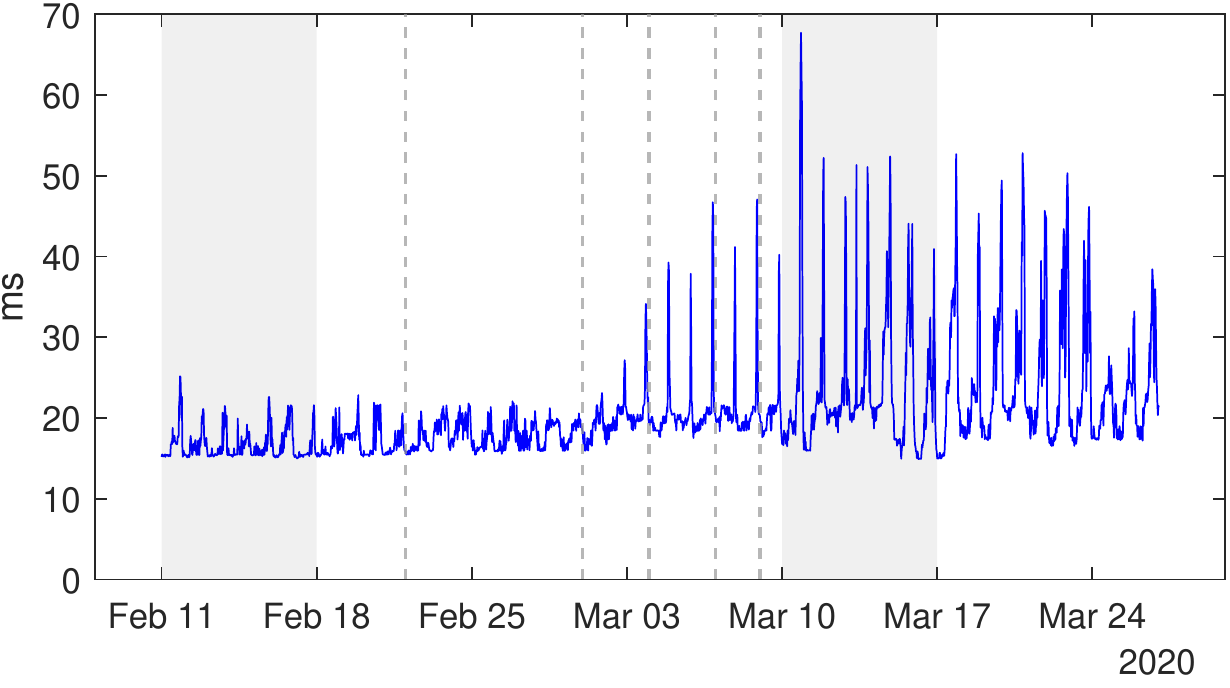}
     }\\
     \caption{Italian AMs, from raw data to aggregated values (scales on the $y$ axis are different).\label{fig:raw_to_aggregated}}
\end{figure}

Figure~\ref{fig:all_data_amd} shows the raw RTT values collected in Italy, AMs, before any processing. More in detail, the scatterplot has been produced using just 1:1000 of the values to make the image readable. The largest values tend to be slightly higher after lockdown events. However, a trend does not clearly emerge from raw values, probably because of the large amount of values and because they are related to paths with very different lengths. The red curve that is also shown in Figure~\ref{fig:all_data_amd} is the moving average of raw AMs values, computed using a window with a duration of five minutes. In this case, increased variability can be observed in the right-hand side of the figure because of lockdown. The period of fluctuations suggests that daily activities may play a role and for this reason they are further analyzed in Section~\ref{sec:circadian}.
Figure~\ref{fig:absolute} shows the absolute values, again for the Italian AMs, but now grouped according to the method defined in Section~\ref{sec:method}. The $d_{min}$ values have been grouped in buckets and averaged, similarly to $r_{min}$, but without subtracting $m$. Let us call such values $s_{min}$. The curve is approximately the same shown in Figure~\ref{fig:all_data_amd}, but with less spikes as the size of buckets is 30 minutes. Figure ~\ref{fig:absolute_and_relative} shows the same $s_{min}$ curve, but now together with $r_{min}$.  The scale is different to have a better view of the phenomenon. The $r_{min}$ curve is very similar to the other one, but translated downwards, as now the $m$ value found for each pair is subtracted.
Figure~\ref{fig:p95} shows the 95th percentile of $q_{min}$, again grouped in buckets of 30 minutes. It is thus computed similarly to $s_{min}$, but now using the 95th percentile instead of the average. Variability of delays is even more visible in this case. 

We do not provide an analogous discussion about the European dataset and/or UDMD for the sake of brevity. 

In the remaining of this paper, we base the analysis mostly on $r$, as we are interested in evaluating the impact of the pandemic on the variable component of latency (with the exception of Section~\ref{subsec:overall} where we also include $s$). The phenomenon is principally observed in its evolution along the line of time, as we wish to understand if, and to what extent, the lockdown measures, enforced in Italy and in the other considered European countries, had an impact on the network.

\section{Impact of lockdown on the Italian Internet}
\label{sec:result}

We studied the impact of the COVID-19 pandemic on the latency of the Italian Internet from different perspectives: when both source and target are located in Italy or just one of the two, when considering the time of the day and workweek/weekend, and when taking into account the version of the Internet Protocol. We also studied the observed latency when the target is part of a content provider network. Beside ICMP-based latency, an evaluation of latency as seen at the HTTP level is provided. Finally, a subsection also shows the path changes occurred at the AS-level as they can be related to the phenomenon under observation.

\subsection{Overall results}
\label{subsec:overall}

\begin{figure*}[t]
    \centering
    \subfloat[][$s$ values, AMD]{
        \label{subfig:ititabsoluteanchoring}
        \includegraphics[width=0.48\textwidth]{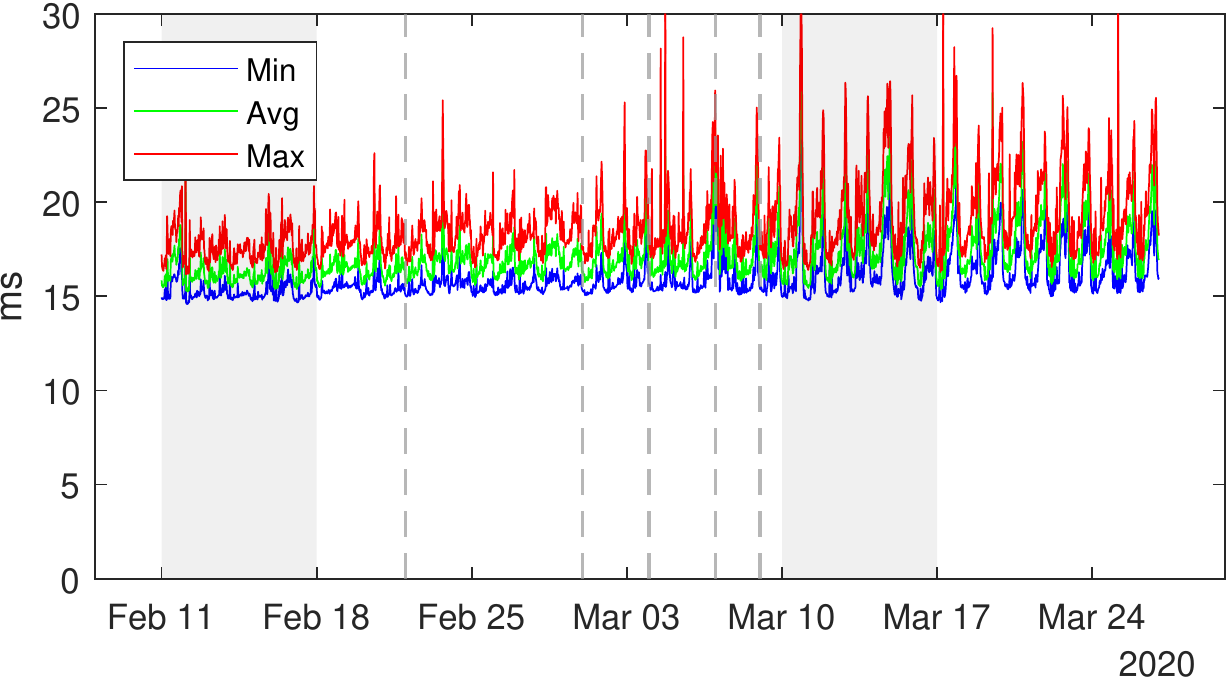}
    }
    \subfloat[][$s$ values, UDMD]{
        \label{subfig:ititabsoluteall}
        \includegraphics[width=0.48\textwidth]{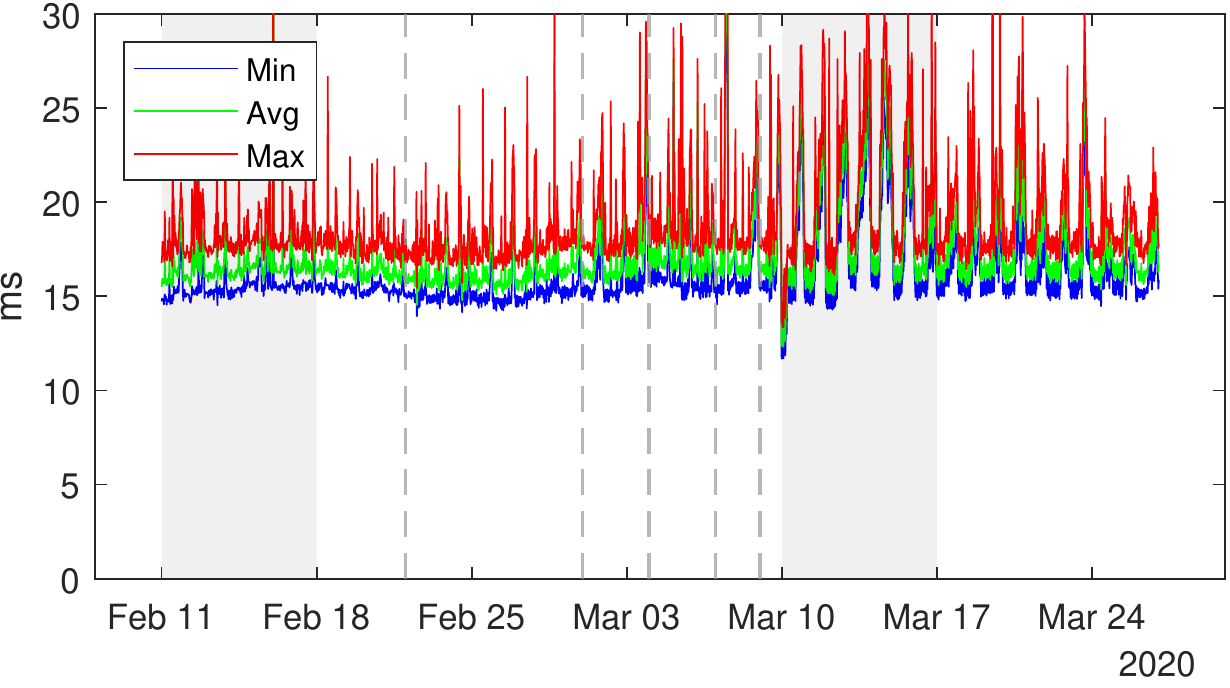}
    }\\
    \subfloat[][$r$ values, AMD]{
        \label{subfig:ititanchoring}
        \includegraphics[width=0.48\textwidth]{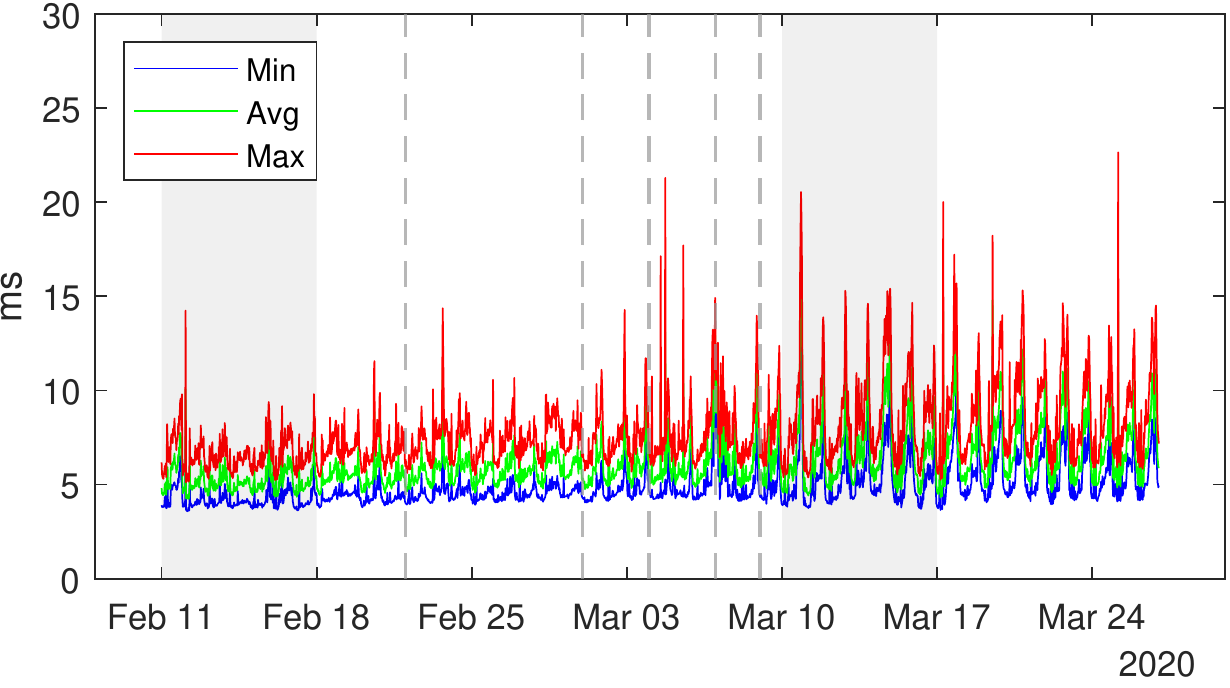}
    }
    \subfloat[][$r$ values, UDMD]{
        \label{subfig:ititall}
        \includegraphics[width=0.48\textwidth]{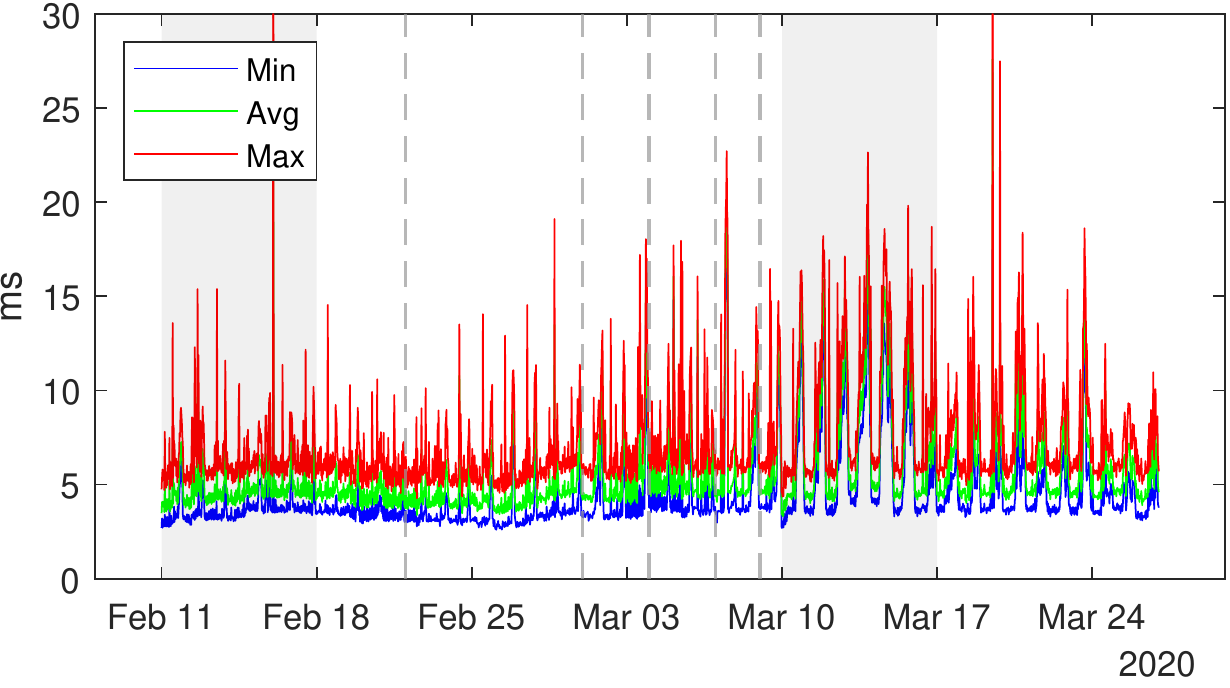}
    }
    \caption{$d$ and $r$ values in measurements with both source and target in Italy. The dashed vertical lines correspond to lockdown events. The two gray areas correspond to W1 and W2.}
    \label{fig:itit}
\end{figure*}

We start our analysis from the measurements in AMD with both source and target in Italy. We first analyze the values of $s$ measured in Italy. We recall that $s$ is the average of the $d$ values in each bucket, as defined in Section~\ref{sec:preliminary}. Figure~\ref{subfig:ititabsoluteanchoring} shows the evolution of $s_{min}$, $s_{avg}$, and $s_{max}$ for the whole observation period. We notice that all values progressively increase over time. Since Italy experienced some partial lockdown events before the most restrictive one, there is no step-like increase, rather a continuous one. However, approximately from February 23, when schools were closed in Northern Italy, delays start to grow and higher peaks can be observed. The largest increase can be observed on March 10, which is the date of the complete Italian lockdown. The increase appears to be quite significant, for example if we consider the $s_{min}$ line, the peaks during the lockdown are approximately 30\% higher than the ones before the lockdown.

Besides the generally increasing value of all the three curves corresponding to the three $s$ variations, the higher variability of latency is also evident. To evaluate the increased variation we compared the value of $s_{min}$ in W1 and W2.
The average $s_{min}$ experiences a 7.6\% increase in W2, and the $s_{min}$ standard deviation experiences a 201.5\% increase. The measurements in UDMD show a similar pattern, but more accentuated, as can be seen in Figure~\ref{subfig:ititabsoluteall}. In fact, in W2, the average $s_{min}$ is 14.5\% higher than in W1, and the $s_{min}$ standard deviation is 310.1\% higher. It has to be noticed that the $s$ values in UDMD show a little drop around March 10, which lasts some hours. We investigated on this aspect, and found that one of the targets of the measurements was unreachable in that interval. The target involved multiple source-target pairs, corresponding to approximately $1/5$ of the measurements, which usually were providing relatively high absolute delays. In other words, this is an artifact due to a temporary lack of measurements, which can happen when measuring real world devices.

Similar patterns can be observed when analyzing the evolution of $r_{min}$, $r_{avg}$, and $r_{max}$ for both AMD and UDMD, which can be found in Figures~\ref{subfig:ititanchoring} and~\ref{subfig:ititall}. In this case however, the curves are translated downwards, as expected. This reflects on the increment of the average $r_{min}$ value in W2 with respect to W1, which is 27.7\% in AMD and 66.8\% in UDMD, but not on the increment of the $r_{min}$ standard deviation value, which remains almost unchanged: 203.6\% in AMD and 280.8\% in UDMD. 

\begin{figure*}[t!]
    \centering
    \subfloat[][AMD, from Italy to the rest of Europe.]{
        \label{subfig:oe-it-all-pp-anchoring-both-all}
        \includegraphics[width=0.48\textwidth]{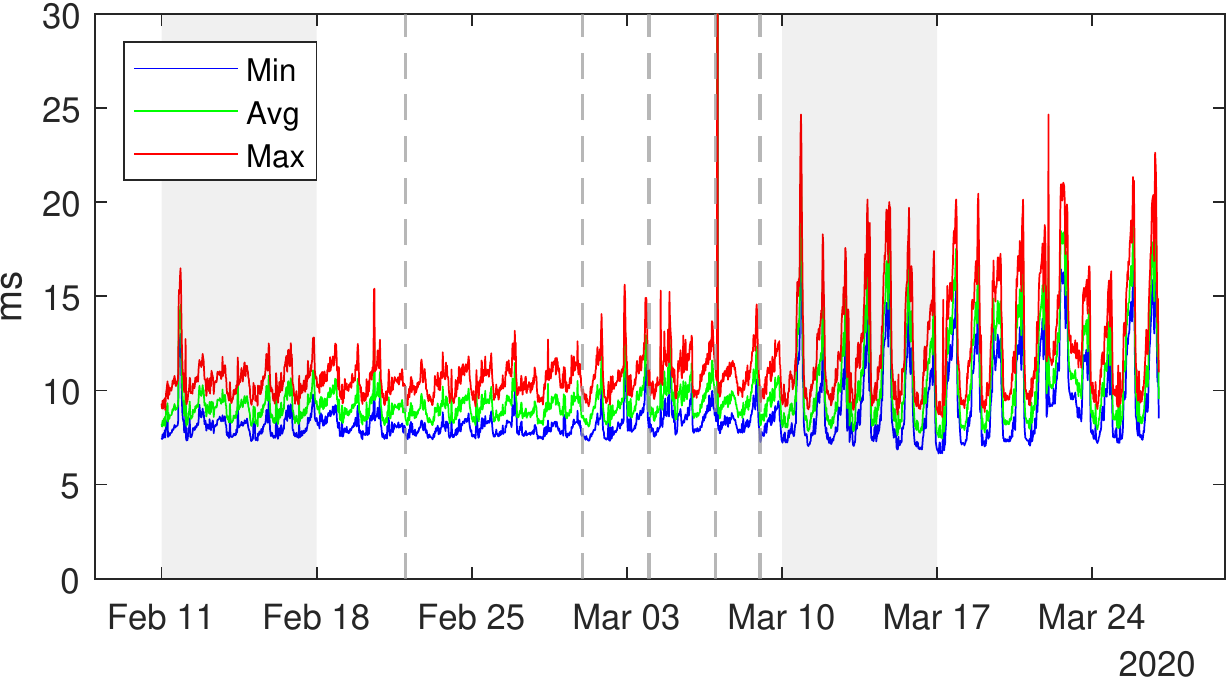}
    }
    \subfloat[][UDMD, from Italy to the rest of Europe.]{
        \label{subfig:oe-it-all-pp-udm-both-all}
        \includegraphics[width=0.48\textwidth]{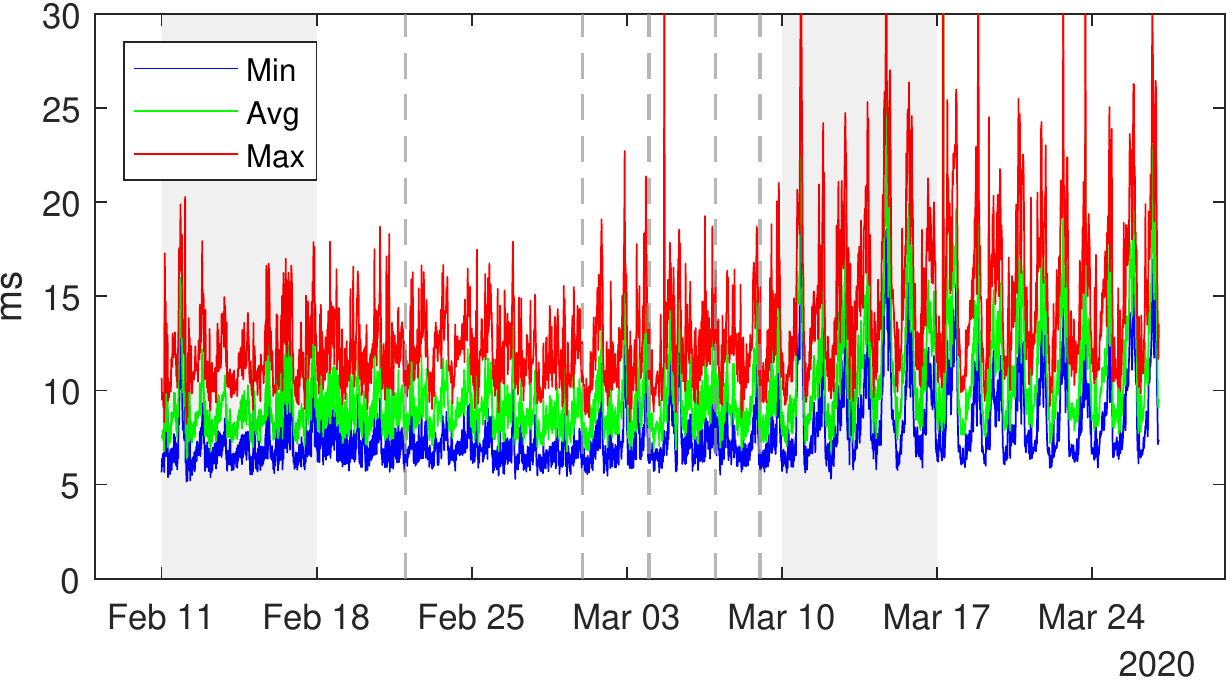}
    }\\
    \subfloat[][AMD, from the rest of Europe to Italy.]{
        \label{subfig:oe-all-it-pp-anchoring-both-all}
        \includegraphics[width=0.48\textwidth]{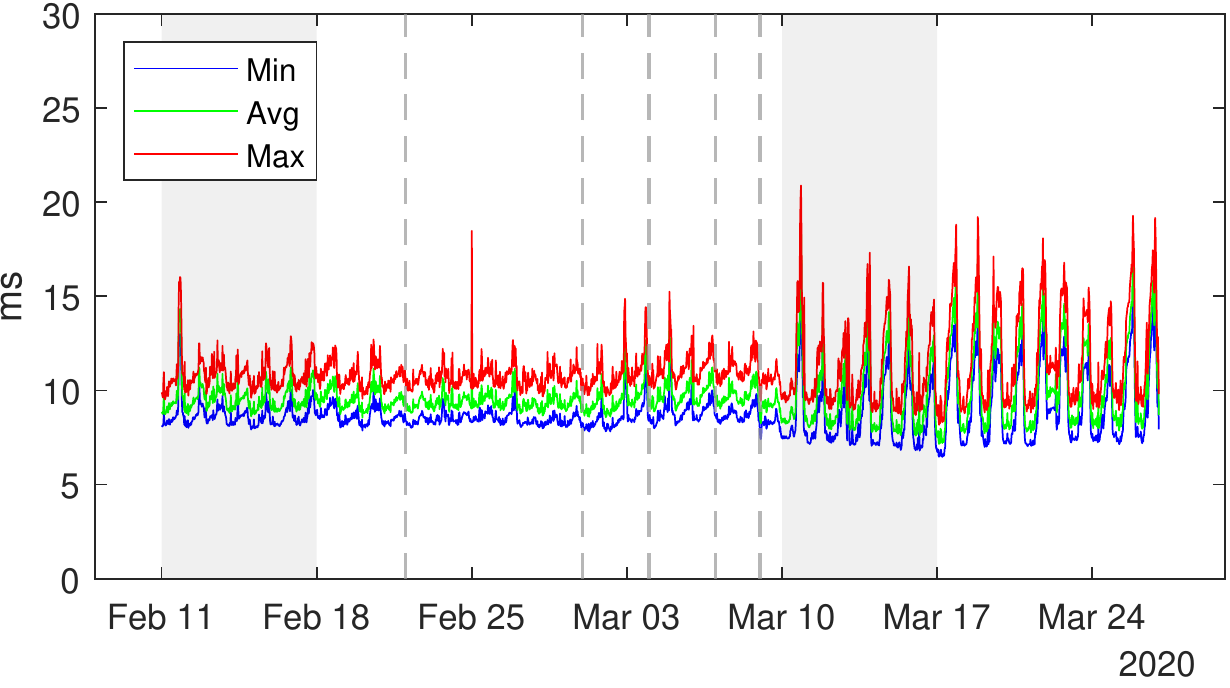}
    }
    \subfloat[][UDMD, from the rest of Europe to Italy.]{
        \label{subfig:oe-all-it-pp-udm-both-all}
        \includegraphics[width=0.48\textwidth]{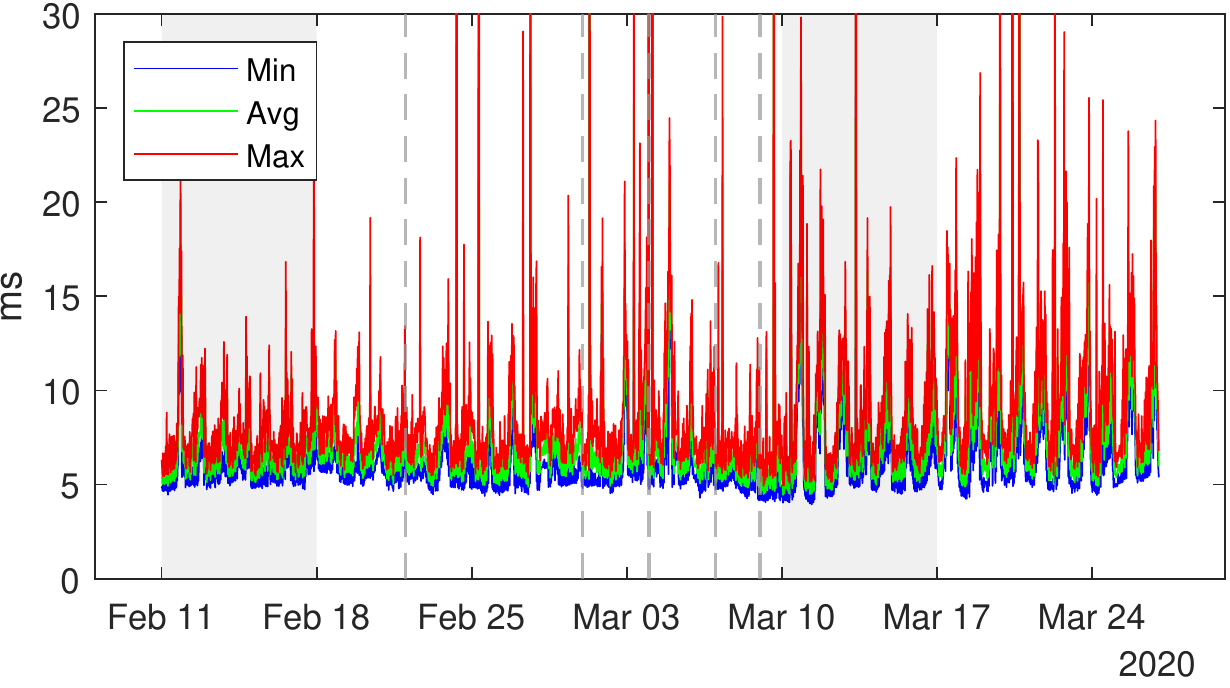}
    }
    \caption{$r$ in measurements from Italy to the rest of Europe and vice-versa. The dashed vertical lines correspond to lockdown events. The two gray areas correspond to W1 and W2.}
    \label{fig:euit}
\end{figure*}

Similar considerations can be made about measurements with sources in Italy and targets spread all over Europe (excluding Italy), as shown in Figures~\ref{subfig:oe-it-all-pp-anchoring-both-all} and~\ref{subfig:oe-it-all-pp-udm-both-all} for AMD and UDMD respectively. Also in this case, the generally increased variability during lockdown is clearly visible.

When considering measurements with sources in Europe and targets in Italy, as shown in Figures~\ref{subfig:oe-all-it-pp-anchoring-both-all} and~\ref{subfig:oe-all-it-pp-udm-both-all}, the situation is a bit different between AMD and UDMD. In particular, for AMD the pattern is approximately the same as in Figure~\ref{subfig:oe-it-all-pp-anchoring-both-all}. For UDMD, instead, the pattern is quite different. An increase of the overall variability is still noticeable, however the three curves appear to be much more squeezed on top of each other, i.e. the $r_{min}$ and $r_{max}$ curves are much closer to each other than in the other scenarios.

It is also worth to notice that, especially for AMD, the local minima of the $r_{min}$ curve tend to get higher during the the transitory period but then they start to get lower. This is particularly visible in Figures~\ref{subfig:ititabsoluteanchoring} and~\ref{subfig:ititanchoring}. The local minima (the troughs) correspond to night hours, when the network is lightly loaded. This phenomenon could be explained by the infrastructural enhancements introduced by network operators to respond to the crisis. For example, during the transition period, TIM (the Italian incumbent) started peering again in public peering LANs of Italian IXPs for the first time since the end of 2012~\cite{tim}. Also, IXPs reported an increase of traffic of 30 - 40\%, which pushed them to introduce upgrades in the capacity of their peering LANs, as reported during the Italian Network Community Meeting held for the occasion~\cite{itam}. Such improvements could also justify the situation in Figures~\ref{subfig:oe-it-all-pp-anchoring-both-all} and~\ref{subfig:oe-all-it-pp-anchoring-both-all}, where the troughs during the lockdown reach smaller values compared to the one of the first analyzed week.

\subsection{Packet loss}
\label{subsec:packetloss}

\begin{figure*}[t!]
    \centering
    \subfloat[][AMD.]{
        \label{subfig:packetloss-europe-europe-it-all-pp-anchoring-both-all}
        \includegraphics[width=0.48\textwidth]{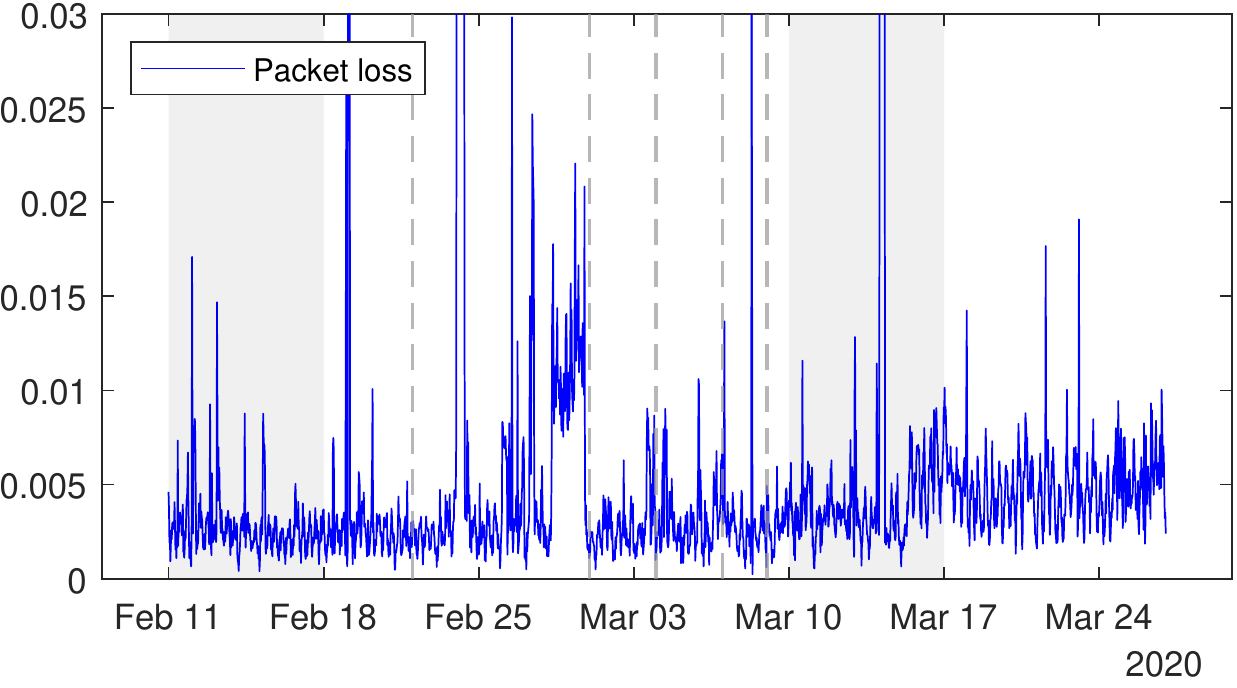}
    }
    \subfloat[][UDMD.]{
        \label{subfig:packetloss-europe-europe-it-all-pp-udm-both-all}
        \includegraphics[width=0.48\textwidth]{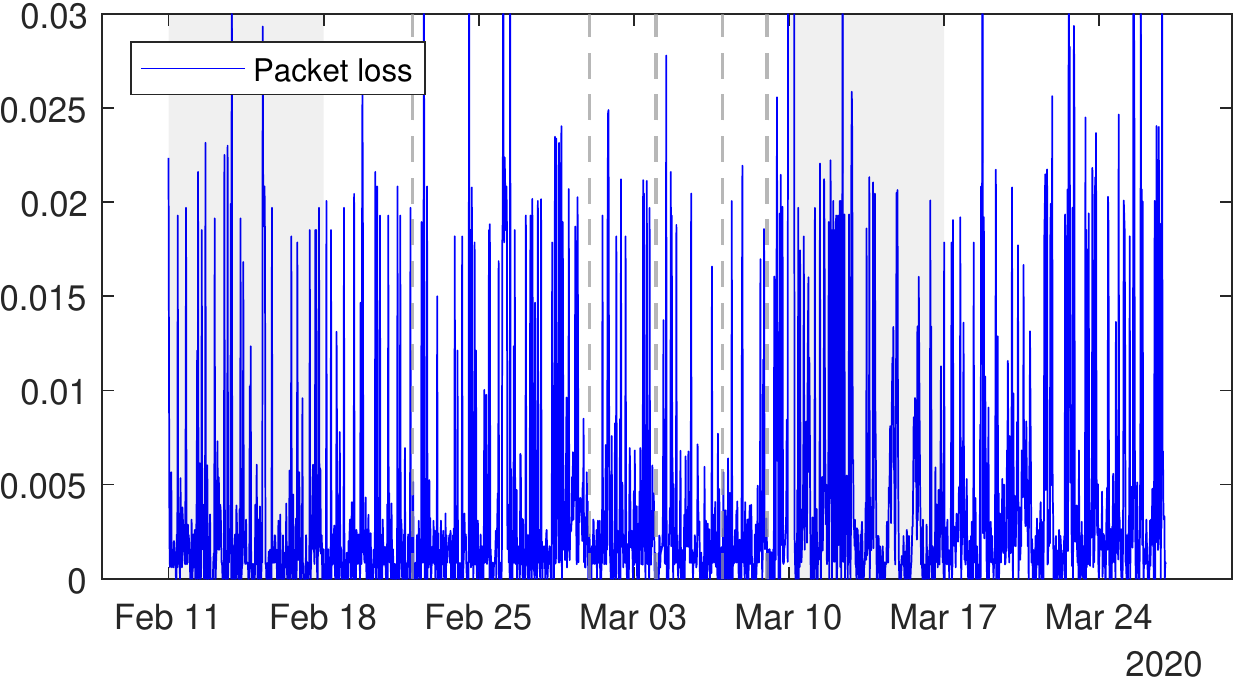}
    }
    \caption{Packet loss in measurements with both source and target in Italy.}
    \label{fig:ploss}
\end{figure*}

We estimated the impact of the lockdown in terms of packet loss, as the fraction of unsuccessful echo request/reply. Figure~\ref{fig:ploss} shows the evolution of the packet loss for the whole observation period, for both AMD and UDMD. Both datasets show an increase of packet loss over time, especially after the lockdown starts. This is consistent with the previous results, and indicates a generally increased congestion due to the lockdown. We computed the average packet loss in W1 and W2. The increase is significant: from 2.8e-3 to 5.9e-3 (+110.3\%) for AMD, and from 3.1e-3 to 9.4e-3 (+205.7\%) for UDMD. Also the standard deviation  of the packet loss rate in W2 increases significantly compared to W1: from 1.8e-3 to 1.1e-2 for AMD (+507.2\%), and from 5.6e-3 to 2.6e-2 for UDMD (+367.9\%).

We notice that in AMD there is an increase of the packet loss that lasts approximately 2 days, around February 29 -- March 1. We investigated this behaviour and found that the increase is not attributable to a narrow subset of the source-target pairs. We did not register any disconnections of sources nor targets, and the same packet loss pattern can be seen in measurements directed to 74\% of the targets, although with different intensities. These targets are geographically spread all over Italy. In particular there are two targets that contribute in a significant manner, which are located in Milan (at the Milan Internet eXchange) and Monopoli (Puglia region), respectively (we point out that measurements towards one target involve multiple source-target pairs). The temporary increase of packet loss is not reflected in the $r$ curve, as for the involved source-target pairs the packets that are correctly delivered experience a negligible increase of the RTT.

\subsection{Circadian rhythms}
\label{sec:circadian}

\begin{figure}[t!]
    \centering
        \includegraphics[width=0.48\textwidth]{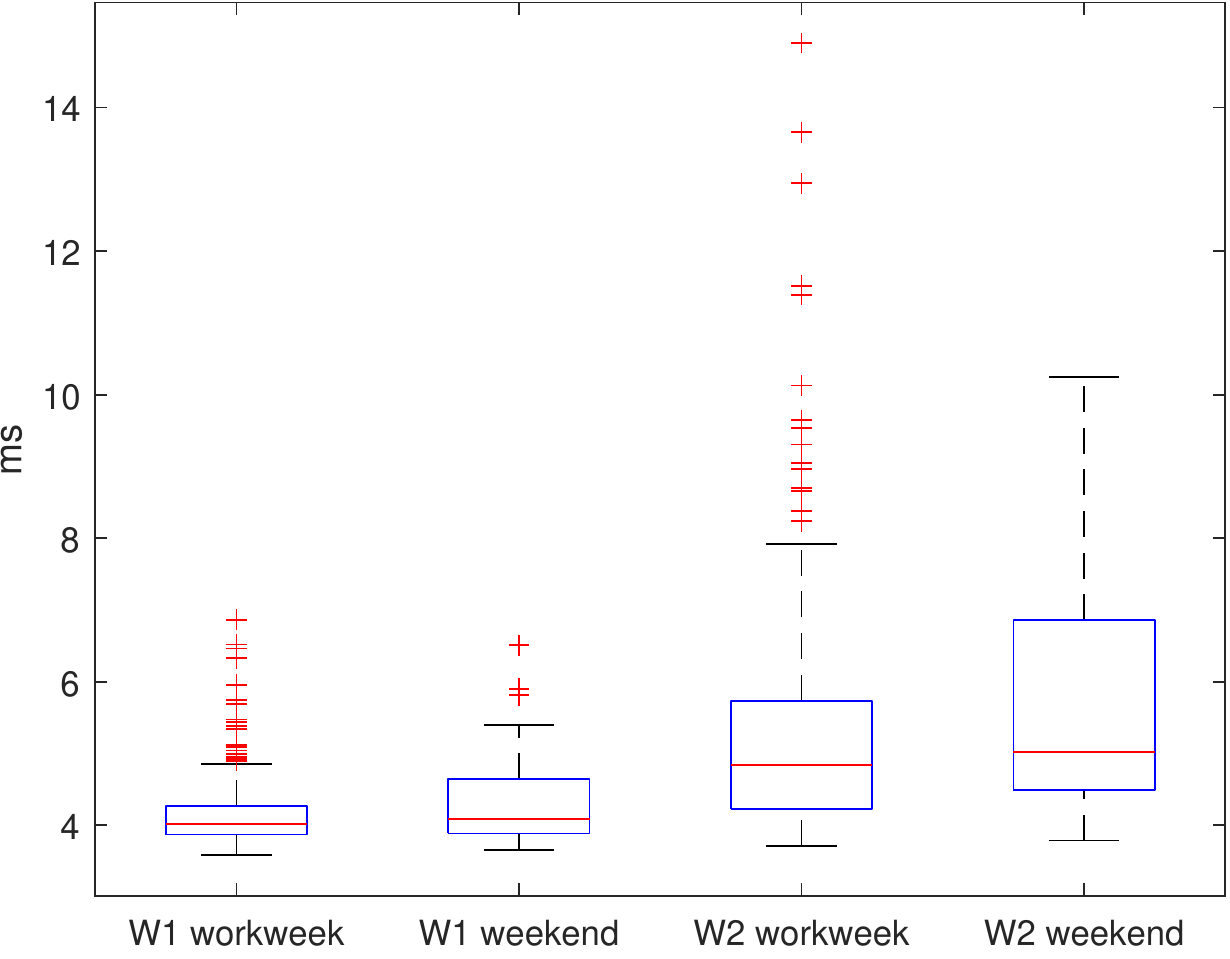}
    \caption{Boxplot of $r_{min}$ in work days and weekend for AMD with both source and target in Italy: comparison between W1 and W2. The red line represents the median value, bottom and top edges indicate the 25th and 75th percentiles, whiskers extend to $\pm2.7$ std. dev., the most external values are depicted using the ``+'' symbol. Note that the y axis scale does not start from $0$.}
    \label{fig:boxplot}
\end{figure}

\begin{figure}[t!]
    \centering
    \subfloat[][AMD, measurements with both source and target in Italy.]{
        \label{subfig:it-it-am-ratio}
        \includegraphics[width=0.48\textwidth]{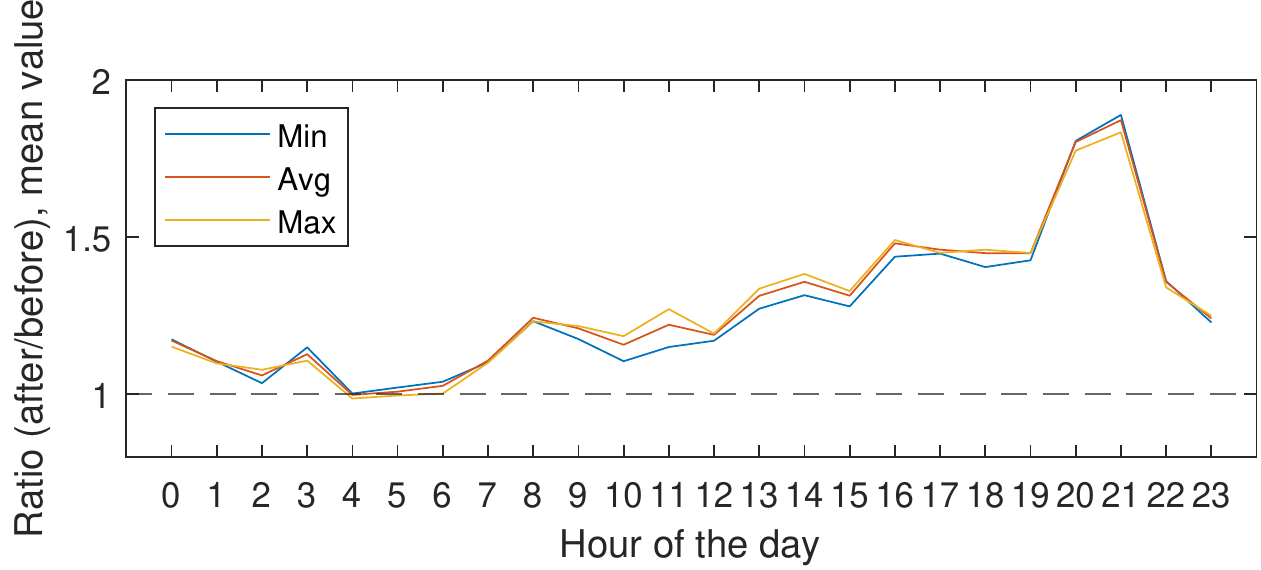}
    }\\
    \subfloat[][AMD, measurements with source in Italy and target in Europe.]{
        \label{subfig:it-all-am-ratio}
        \includegraphics[width=0.48\textwidth]{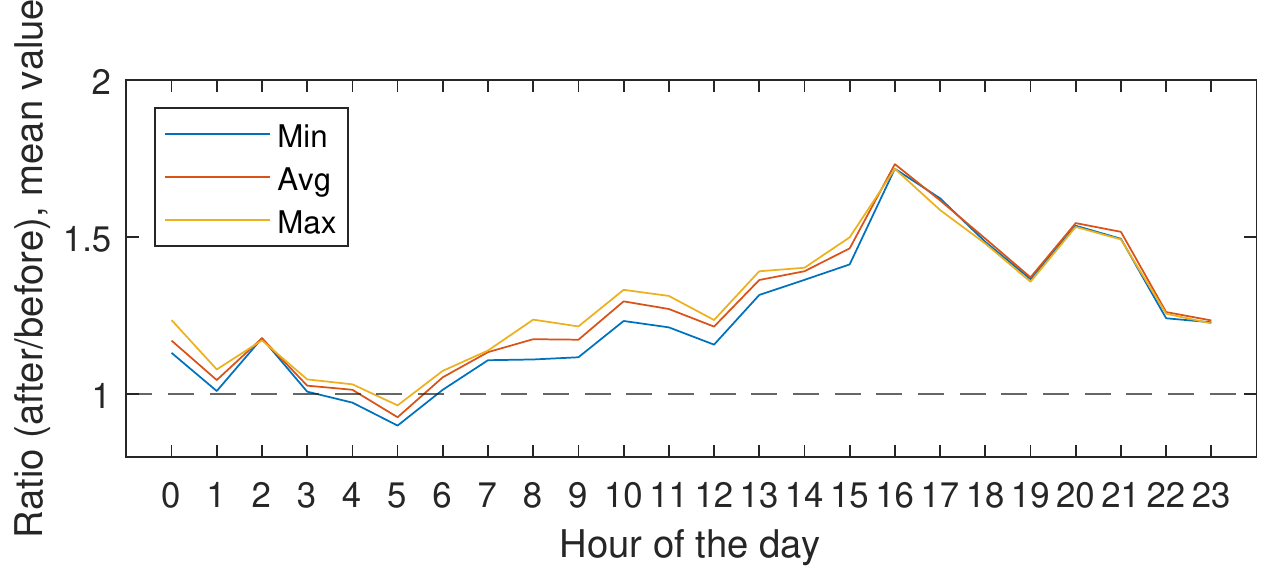}
    }\\
    \subfloat[][AMD, measurements with source in Europe and target in Italy.]{
        \label{subfig:all-it-am-ratio}
        \includegraphics[width=0.48\textwidth]{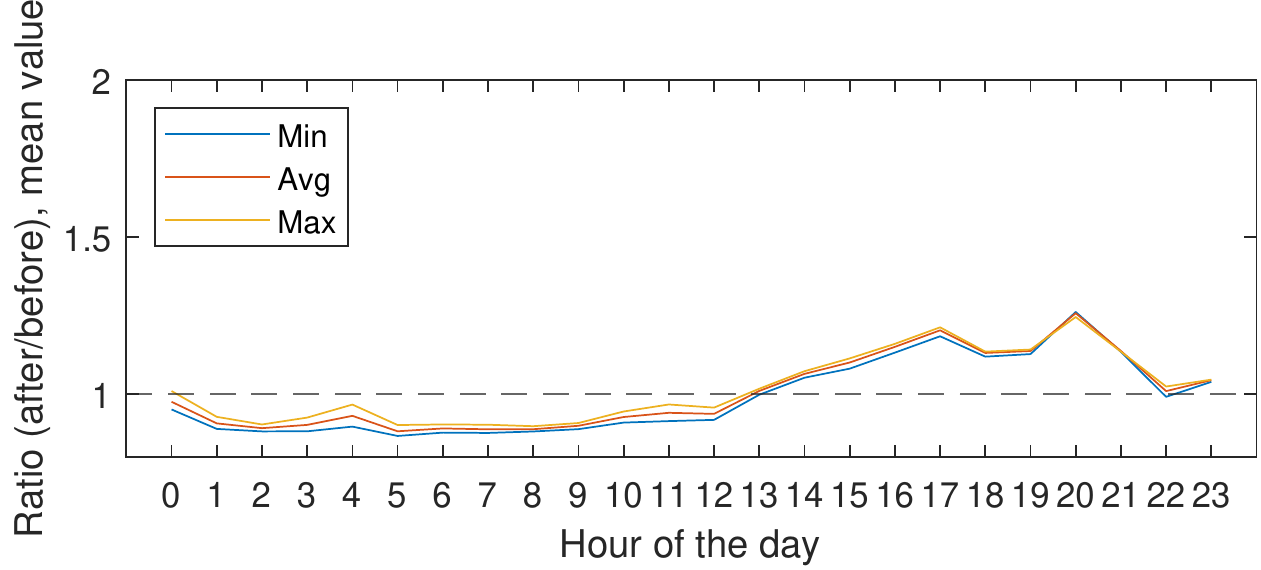}
    }
    \caption{Ratio of $r$ in W2 to $r$ in W1, for all hour slots. Hours are expressed in UTC time (Italy is UTC+1 for the observed period).}
    \label{fig:it-it-ratio}
\end{figure}

\begin{figure*}[t]
     \centering
     \subfloat[][Peak hours.]{
         \label{subfig:cdf-peak}
         \includegraphics[width=0.4\textwidth]{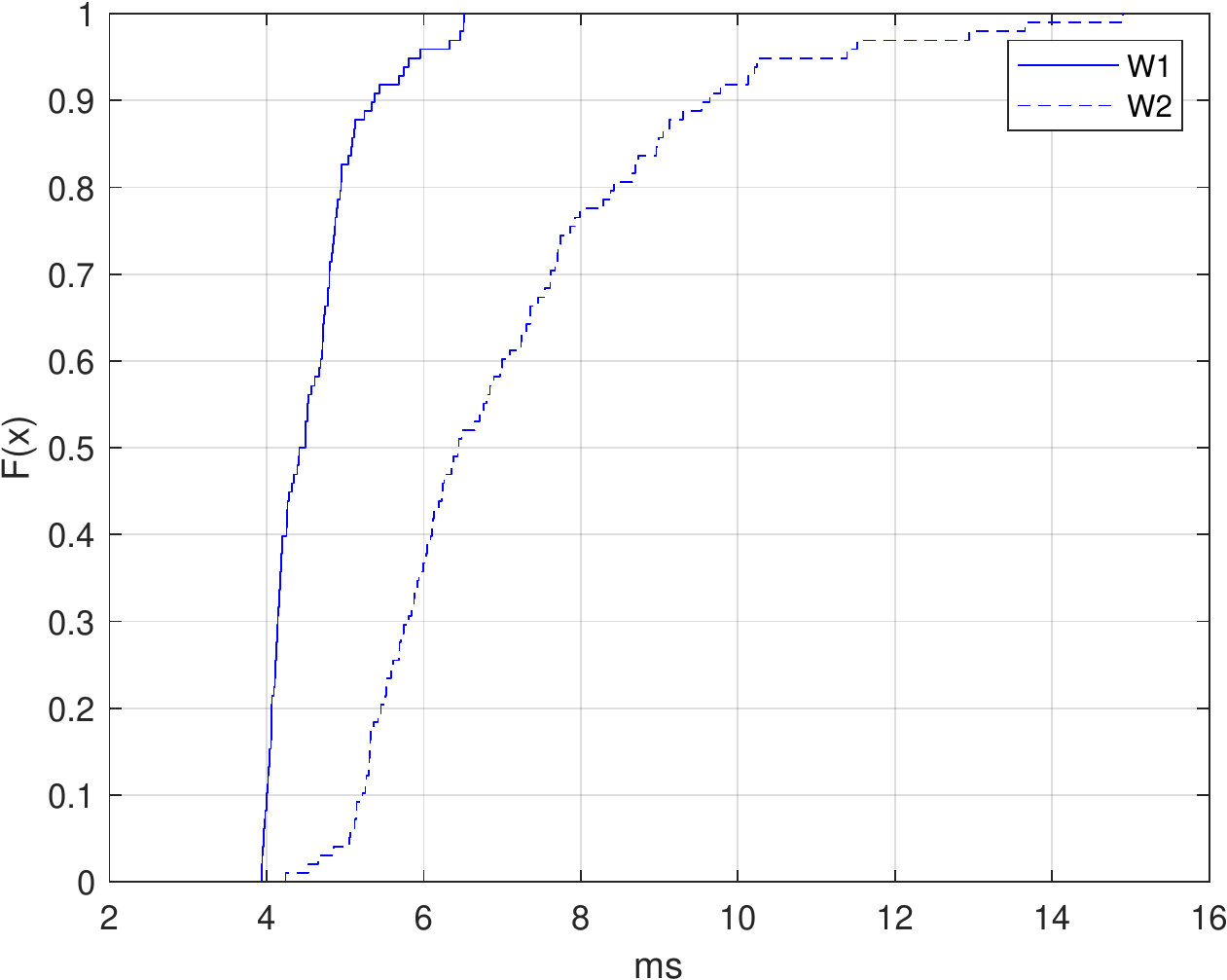}    
    }
     \subfloat[][Off-peak hours.]{
         \label{subfig:cdf-offpeak}
        \includegraphics[width=0.4\textwidth]{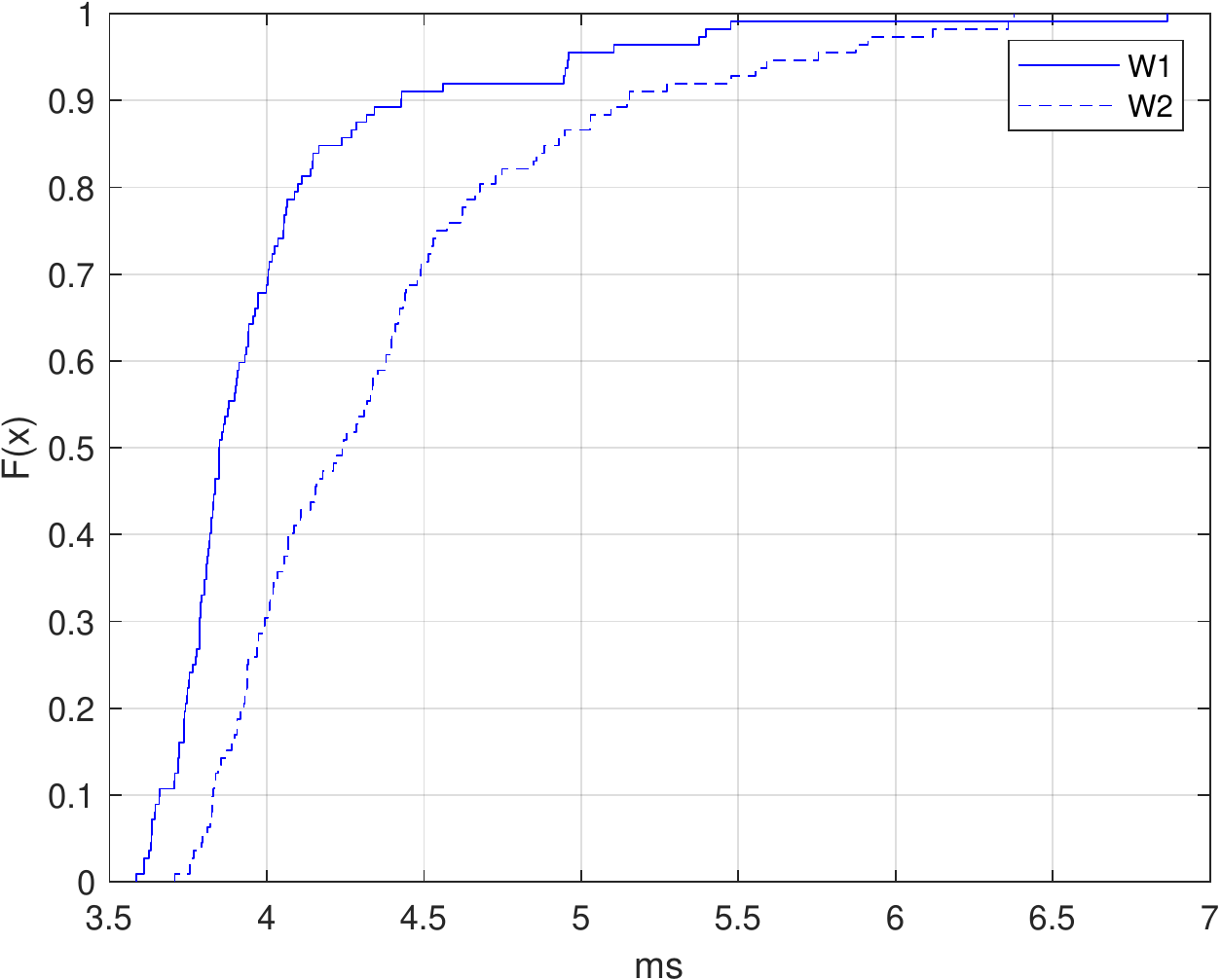}
    }
    \caption{Peak hours vs off-peak hours, $r_{min}$ values in W1 and W2 for AMD measurements with source and target in Italy.}
    \label{fig:peakoffpeak}
\end{figure*}

The three curves shown in Figures~\ref{fig:itit} and~\ref{fig:euit} show repeated peaks and troughs. This is particularly evident during lockdown, and suggests that latency gets more influenced by weekly or circadian rhythms.

To evaluate the influence of the day of the week, we separated the Italian measurements in those run during work days and those run during weekends for W1 and W2. Results can be seen in Figure~\ref{fig:boxplot}, which shows the boxplot of $r_{min}$ in W1 and W2 for work days and weekends. In both W1 and W2 the weekend values are slightly higher and show higher variability. In addition, a significant increase can be observed in W2 compared to W1.

To evaluate the influence of the time of the day on latency, we divided Italian measurements in one-hour slots and aggregated them when executed in the same hour. Then, we calculated for each slot the ratio between the average $r$ values collected during W1 and during W2. Results are represented in Figure~\ref{subfig:it-it-am-ratio} for AMD. It is clearly visible how the increase is not uniform across the time of the day. 
Night hours show no considerable increase. This is not surprising, as human activity is very limited at night, thus also the congestion of the Internet. Morning hours show some little increase. Afternoon hours show a more evident increase, but the highest increase occurs between 16:00 UTC and 22:00 UTC, with a peak between 20:00 and 21:00 UTC.
This is interesting, as, combined with the results of weekend highlighted in Figure~\ref{fig:boxplot}, it suggests that remote working and distance learning have some impact on the Italian Internet latency, but the major effects can be attributed to leisure activities which typically occur in the afternoon/evening and on weekends, such as gaming or video streaming (the reader has to keep in mind that Italy is UTC+1 in the analyzed period). This could be due to the lack of other recreational activities during the lockdown.
To deepen on this aspect we computed the empirical CDFs of the $r_{min}$ values in peak and off-peak hours for W1 and W2 (Figures~\ref{subfig:cdf-peak} and~\ref{subfig:cdf-offpeak}). For peak hours, we considered the interval 16:00-23:00, for off-peak we considered the interval 00:00-08:00. We can notice that in both peak and off peak hours the CDFs for W1 and W2 are clearly separated, and the CDF for W2 shows the highest values. However, it is clearly visible that while in off-peak hours the two CDFs are very similar, in peak hours the CDF for W2 shows a substantial increase of the $r_{min}$ values (note that the x axis scale in the two figures is different).

In Section~\ref{subsec:overall}, we highlighted as in night hours the values of $r_{min}$ are particularly low during the lockdown, and in particular in Figures~\ref{subfig:oe-it-all-pp-anchoring-both-all} and~\ref{subfig:oe-all-it-pp-anchoring-both-all} the local minima during the lockdown seem lower than the ones before the lockdown. To evaluate this effect we computed the ratio between W1 and W2 for the two cases depicted in the aforementioned figures. The results are shown in Figures~\ref{subfig:it-all-am-ratio} and~\ref{subfig:all-it-am-ratio}, respectively. In both cases the ratio goes slightly below one during night and morning hours. Especially surprising is Figure~\ref{subfig:all-it-am-ratio} where the the ratio goes below one for the entire morning. This could happen as in Figure~\ref{subfig:all-it-am-ratio} we consider sources outside Italy, thus the access network is still not involved in a lockdown phase, and targets in the Italian infrastructure, which as mentioned has been improved to cope with the traffic increase. During night and morning hours the load on the network is still light, so in this particular configuration the performance could increase. To conclude, in evening hours the increased Internet usage during lockdown generally produces larger delays, but in  periods of lighter load the network is sometimes more efficient than before the lockdown.

\subsection{IPv4 and IPv6}
\label{subsection:v4v6}

\begin{figure*}[t!]
    \centering
    \subfloat[][AMD, IPv4.]{
        \label{subfig:it-it-v4-am}
        \includegraphics[width=0.48\textwidth]{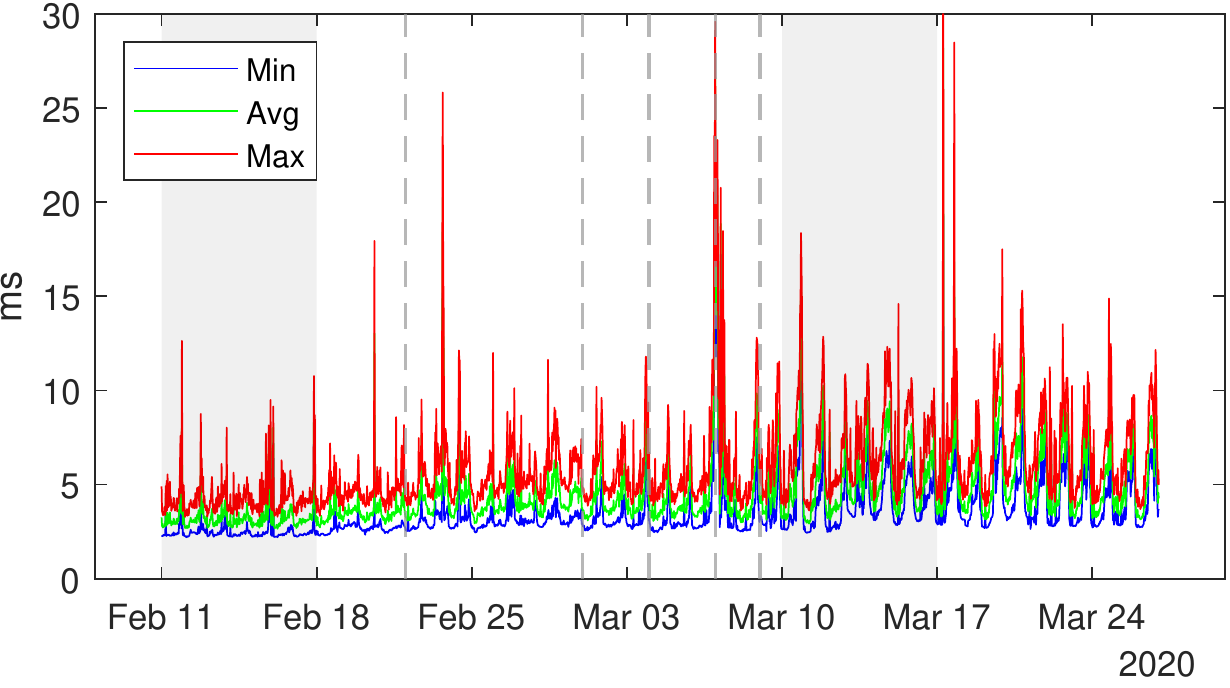}
    }
    \subfloat[][AMD, IPv6.]{
        \label{subfig:it-it-v6-am}
        \includegraphics[width=0.48\textwidth]{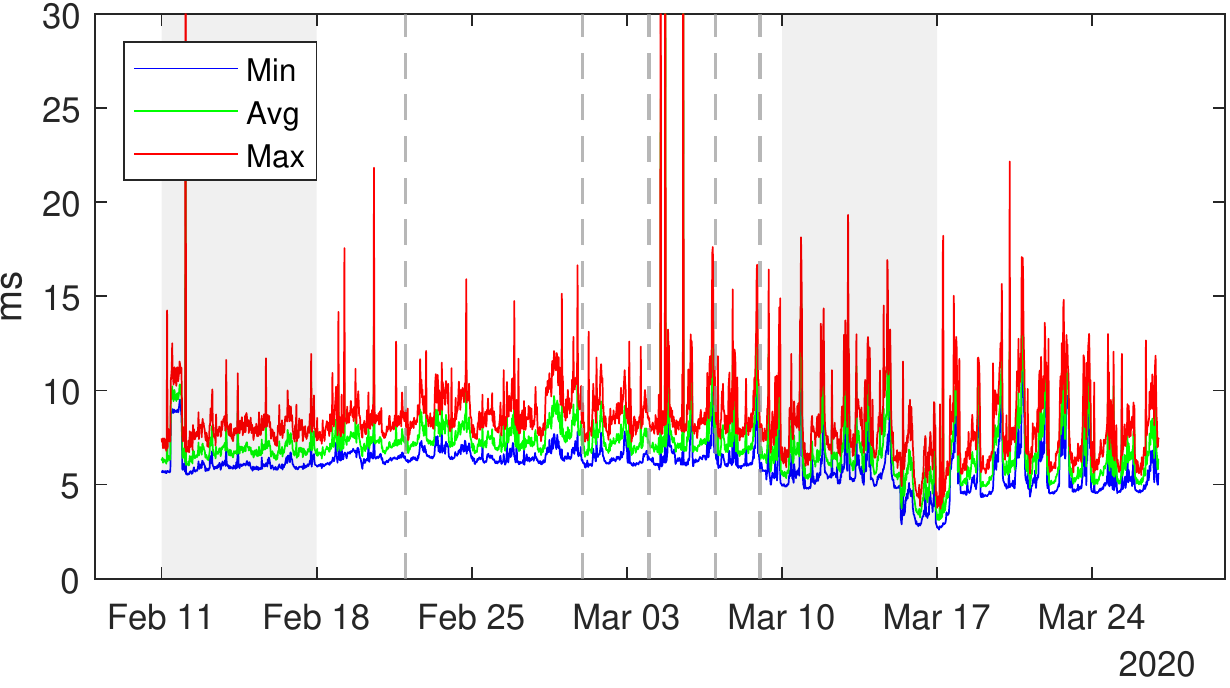}
    }
    \caption{$r$ in measurements with both source and target in Italy: IP version comparison.}
    \label{fig:v4v6_1}
\end{figure*}

We further analyzed AMD taking into account the version of the IP protocol. To perform a fair comparison we only selected measurements run by dual stack probes (i.e., probes which have both IPv4 and IPv6 connectivity). This means that for both IPv4 and IPv6 we consider measurements run between the same sources and targets. We found $11.6$ million IPv4 RTT values and $9.8$ million IPv6 RTT values with both source and target in Italy. Figures~\ref{subfig:it-it-v4-am} and~\ref{subfig:it-it-v6-am} shows the three $r$ variations for IPv4 and IPv6, respectively.
IPv6 latency, in Italy, seems to be characterized by larger variability compared to IPv4 latency, independently from the lockdown period. Such variability increases even more during the lockdown period. However, for IPv6 a reduction of minimum $r_{min}$ values is particularly evident, probably due to the network improvements introduced by operators. In particular, a significant drop around March 17 is visible. We investigated on this aspect and found that a subset of measurements were initially originated by source-target pairs which shared a common upstream provider and were flowing through non national paths (mainly via Germany, Switzerland, and Netherlands). In correspondence of the observed drop, the measurements involving these source-target pairs start to flow through local paths via other providers. These paths show considerably smaller latencies. After few days, the original configuration is restored.

We further studied the latency experienced between 20:00 UTC and 21:00 UTC (the peak hour previously identified) by the two protocol versions. In particular, we compared the values collected during W1 with the values during W2. For IPv4, values of the $r_{min}$ average show a 121.7\% increase in W2. For IPv6 instead, the increase is more modest, 11.2\%. Both show a similar increase in variability in W2: 103.8\% IPv4, and 128.1\% for IPv6.

To conclude, IPv6 in Italy is characterized by a generally higher variability than IPv4, but in peak hours the former has been impacted less than the latter by stay-at-home orders. This is not surprising, as IPv4 and IPv6 are generally served by different infrastructures, and follow different paths~\cite{Dhamdere12:measuring}. In addition, IPv6 is not as common as IPv4 in domestic connectivity, and this could justify the minor impact of the lockdown on latencies observed in IPv6 measurements.

The same analysis was not repeated using UDMD, as the relatively limited amount of IPv6-based UDMs does not allow us to produce statistically sound results.

\subsection{A content delivery example: YouTube}

\begin{figure}[t]
    \centering
    \includegraphics[width=0.48\textwidth]{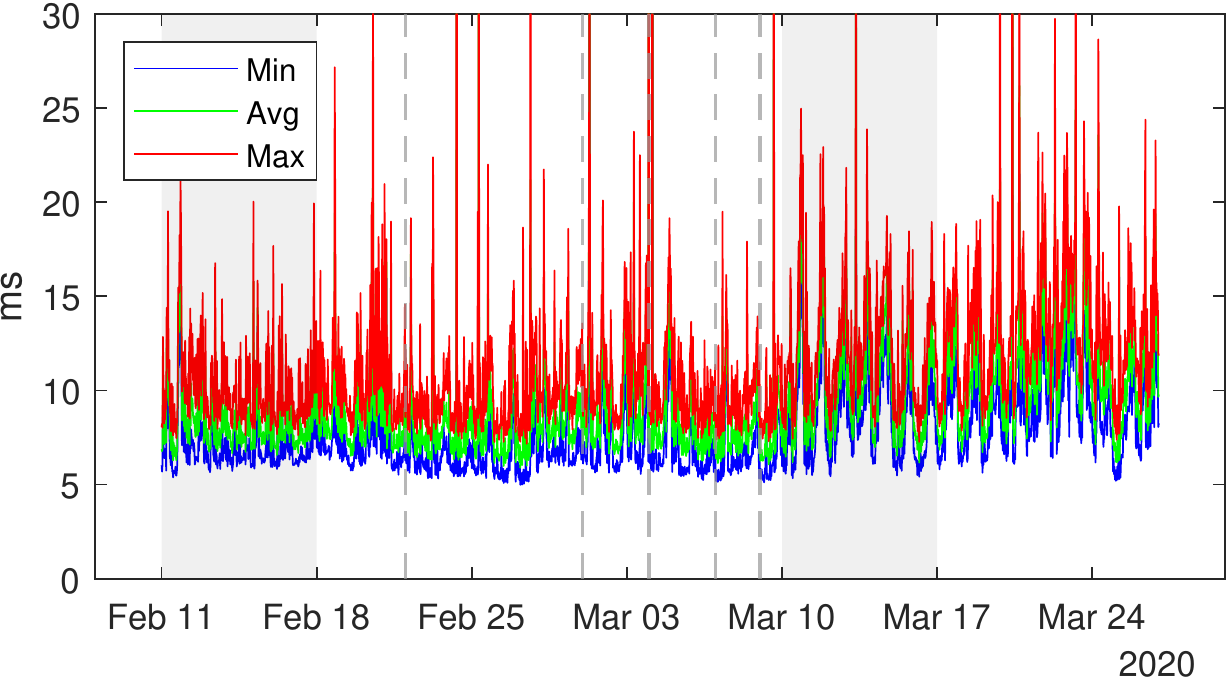}
    \caption{$r$ in measurements towards YouTube servers located in Italy.}
    \label{fig:youtube}
\end{figure}

Since a large fraction of traffic is nowadays directed towards content providers, which we suspect also being related to most of the evening traffic (e.g. video entertainment), we investigated the impact of the lockdown on the latencies towards YouTube.

We collected 
measurements towards the YouTube Content Delivery Networks (CDNs), which is used to serve video content. 
YouTube operates elaborated server selection strategies~\cite{torres2011dissecting} which could lead to inaccurate results. To avoid this, we operated as follows:
\begin{enumerate*}[label=(\roman*)]
    \item We first mapped the names associated with the ad-hoc YouTube CDNs (googlevideos.com and ggpht.com) to the IP addresses that are used to serve content in Italy. For this purpose, we used all the RIPE Atlas probes in Italy to run multiple DNS queries in order to obtain the IP addresses associated with the YouTube servers.
    \item After obtaining all the visible IP addresses of the YouTube's servers (413 addresses), we used the anycast detection service offered by RIPE IPmap, which uses active measurements from RIPE Atlas probes to detect if an IP is anycast or not.
    \item Then, we used again RIPE IPmap to geolocate the IP addresses of the servers. The servers not in Italy were discarded.
    \item Finally, we used these IP addresses to extract RIPE Atlas measurements targeting them. We only selected ICMP ping measurements (which are not subject to HTTP or application layer redirects). We found approximately $6$ million RTT values towards $115$ YouTube's servers located in Italy.
\end{enumerate*}

Figure~\ref{fig:youtube} shows the three $r$ variations we obtained. An increase of the overall latency and its variability is visible during the days of the lockdown. The standard deviation of $r_{min}$ is approximately 86.9\% higher in W2 with respect to W1. The average $r_{min}$ instead increases by 25.2\%.

\begin{figure*}[t]
    \centering
    \subfloat[][RTT values collected between 05:00 UTC and 06:00 UTC]{
        \label{subfig:google-morning}
        \includegraphics[width=0.48\textwidth]{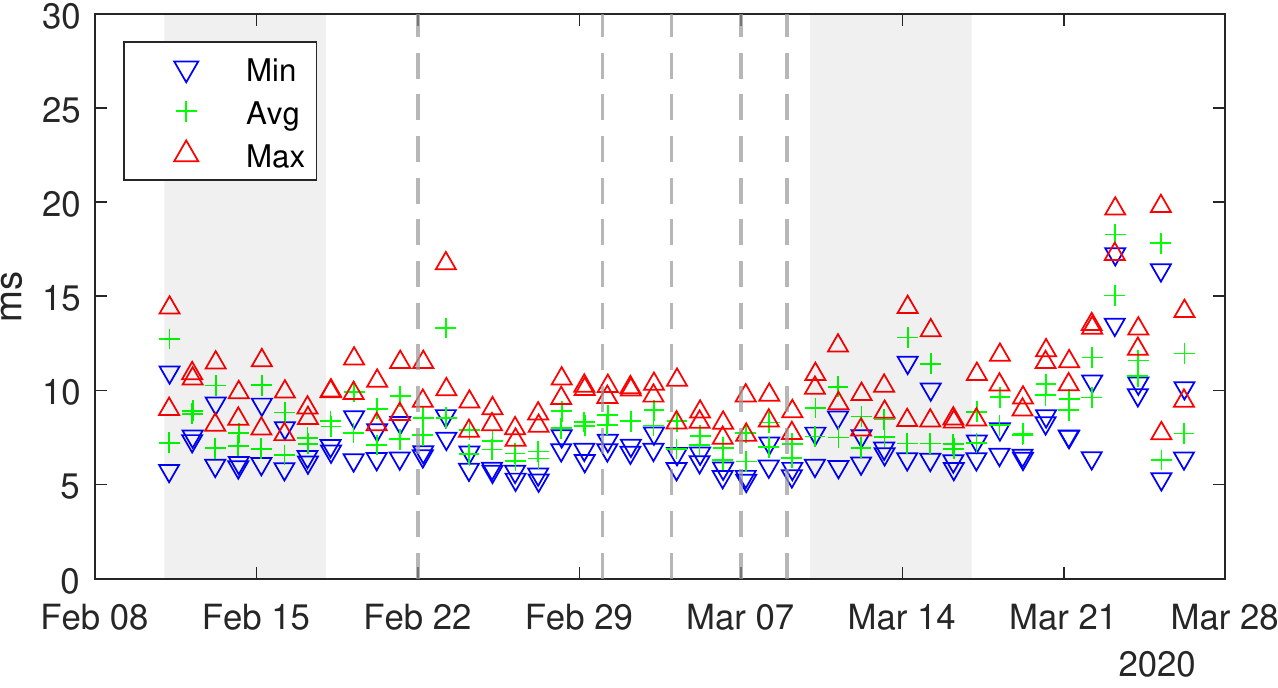}
    }
    \subfloat[][RTT values collected between 20:00 UTC and 21:00 UTC]{
        \label{subfig:google-evening}
        \includegraphics[width=0.48\textwidth]{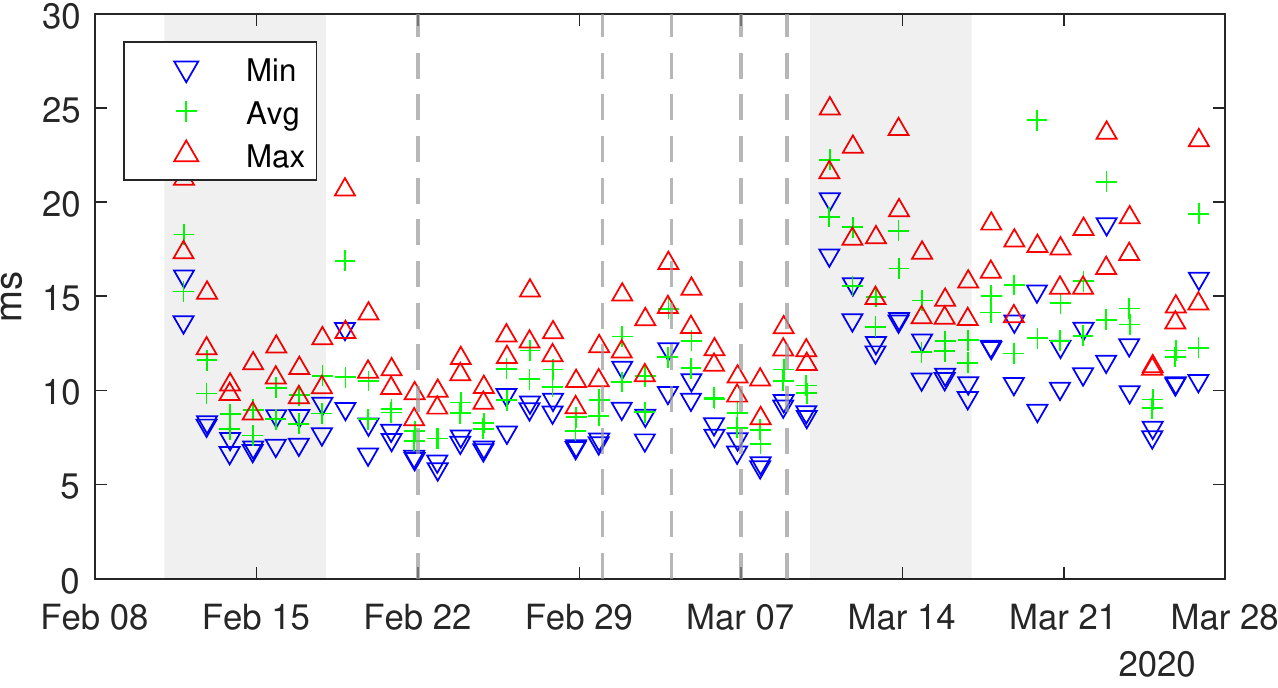}
    }
    \caption{$r$ in measurements towards YouTube servers located in Italy: time slot comparison.}
    \label{fig:cdn_hours}
\end{figure*}

Our hypothesis that the increase of RTT registered in the evening hours is due to people forced to stay home and using the Internet for entertainment, is thus strengthened by the results of the measurements towards YouTube. Figure~\ref{subfig:google-morning} shows measurements collected between 05:00 UTC and 06:00 UTC while Figure~\ref{subfig:google-evening} shows measurements collected between 20:00 UTC and 21:00 UTC, for the entire observation period. 
Also in this case, the $r_{min}$ during night hours slightly improves during the lockdown. The $r_{min}$ in the evening gets moderately higher during the transitory period and abruptly increases after the first day of complete lockdown.

We acknowledge that YouTube is not fully representative of all content providers networks, in fact our initial purpose was to collect measurements towards Facebook and Netflix as well. However, we did not find in RIPE Atlas enough measurements towards Facebook and Netflix servers to cover the whole observation interval and obtain statistically significant results.

\subsection{Path changes}

\begin{figure}[t]
    \centering
    \includegraphics[width=0.48\textwidth]{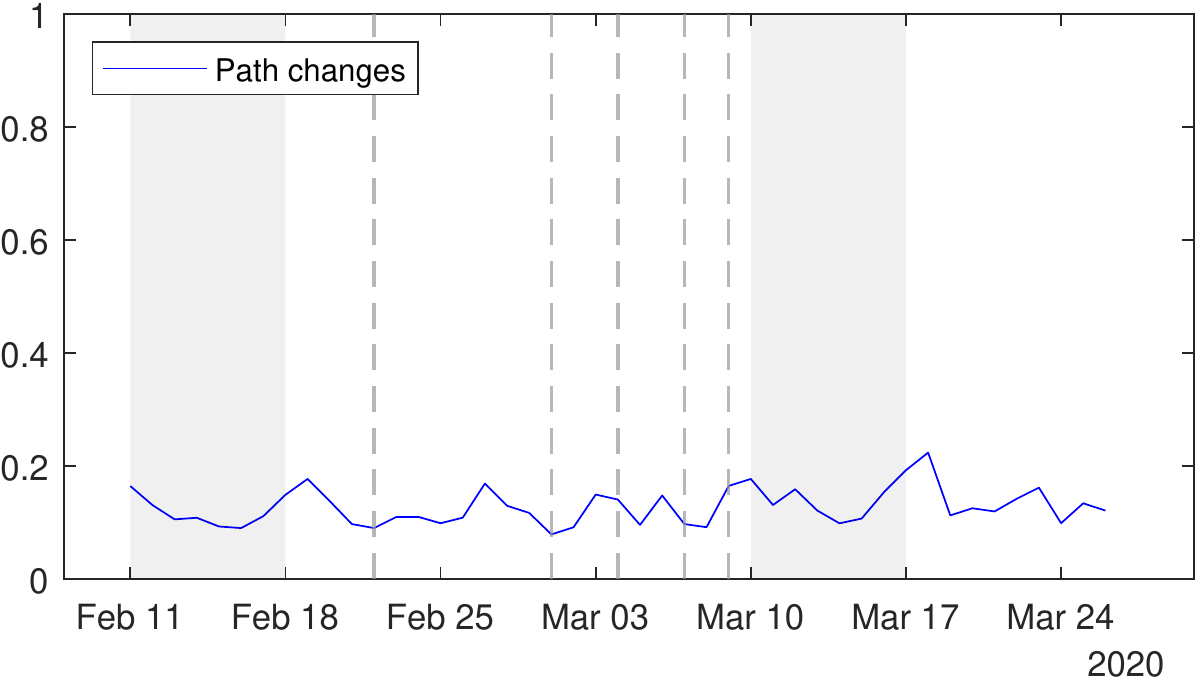}
    \caption{Fraction of the source-target pairs that experience at least one path change at the AS-level during each single day, AMD.}
    \label{fig:path_changes}
\end{figure}

To further improve our analysis, we looked for anomalous patterns in the level of path changes between the sources and targets of our measurements. We consider path changes at the Autonomous System (AS) level, as such level is the one impacting the most on the geography of the paths, while the IP level is subject to various known artifacts and intra-AS load balancing~\cite{Augustin06:avoiding}. From RIPE Atlas, we collected ICMP traceroute measurements to match the Italian source-target pairs of our analysis. For AMD, this was rather straightforward as, besides ICMP ping measurements, AMs also include other types of measurements such as ICMP traceroutes. For UDMD, instead, we could not find enough traceroute measurements to match the source-target pairs of the ping measurements, thus we restricted our analysis to the AMD source-target pairs.

To extract the AS paths from the traceroute IP paths, we use the following methodology:
Initially, we match all the IP addresses against the Internet Topology Data Kit~\cite{itdk} produced by CAIDA. This step converts different IPs belonging to the same router to a single one. Then, we remove the first IP hop inside the network of the probe's host, as it is usually fixed. This step allows us to avoid most of the artifacts introduced by the ICMP rate limiting happening close to the probes~\cite{iodice2019periodic}. If an IP path includes series of wildcards (i.e., non responding hops), these are squashed into a single one. Then, each IP is converted to the AS number originating the prefix such IP belongs to. For this step we use the longest-prefix matching on RIPE RIS data~\cite{ris}. Some addresses cannot be mapped to an AS because they are not available in the RIPE RIS data (e.g. not announced publicly), or, more frequently, they are part of the private address space (RFC1918)~\cite{rekhter1996rfc1918}. We discard such unmappable addresses. This approach has been previously described by Hyun et al.~\cite{hyun2003traceroute}.
More sophisticated approaches are of difficult application due to the size of our dataset. Moreover, our main interest is to quantitatively estimate the variation of the path changes over time, and not to analyze the Internet AS-level topology. 

We then divided the observation period in buckets of one day each, and for each bucket we computed the fraction of source-target pairs that incurred in at least one change of AS path in that bucket. Figure~\ref{fig:path_changes} shows the fraction of path changes over time for AMD. The fraction of source-target pairs experiencing path changes each day is rather uniform on all the observation period, between 0.1 and 0.2, which means that each day just 10-20\% of source-target pairs experienced a change of AS path. However, by comparing W1 and W2 from a numerical point of view, we found that the average fraction of path changes per day is 17.9\% higher in W2.
Thus, there seems to be a slight increase of the number of path changes per day due to the impact of lockdown.

\subsection{HTTP}

\begin{figure*}[t]
    \centering
    \subfloat[][$r$ values.]{
        \label{subfig:httprtt}
        \includegraphics[width=0.48\textwidth]{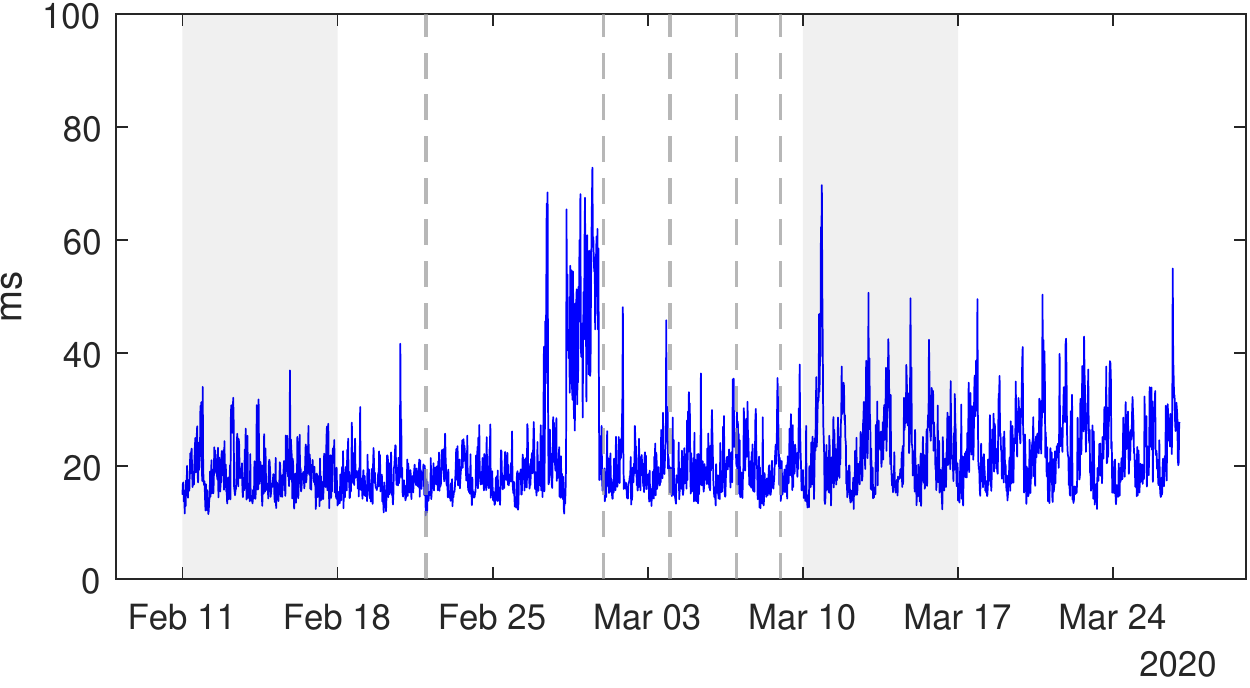}
    }
    \subfloat[][Failure rate.]{
        \label{subfig:httpploss}
        \includegraphics[width=0.48\textwidth]{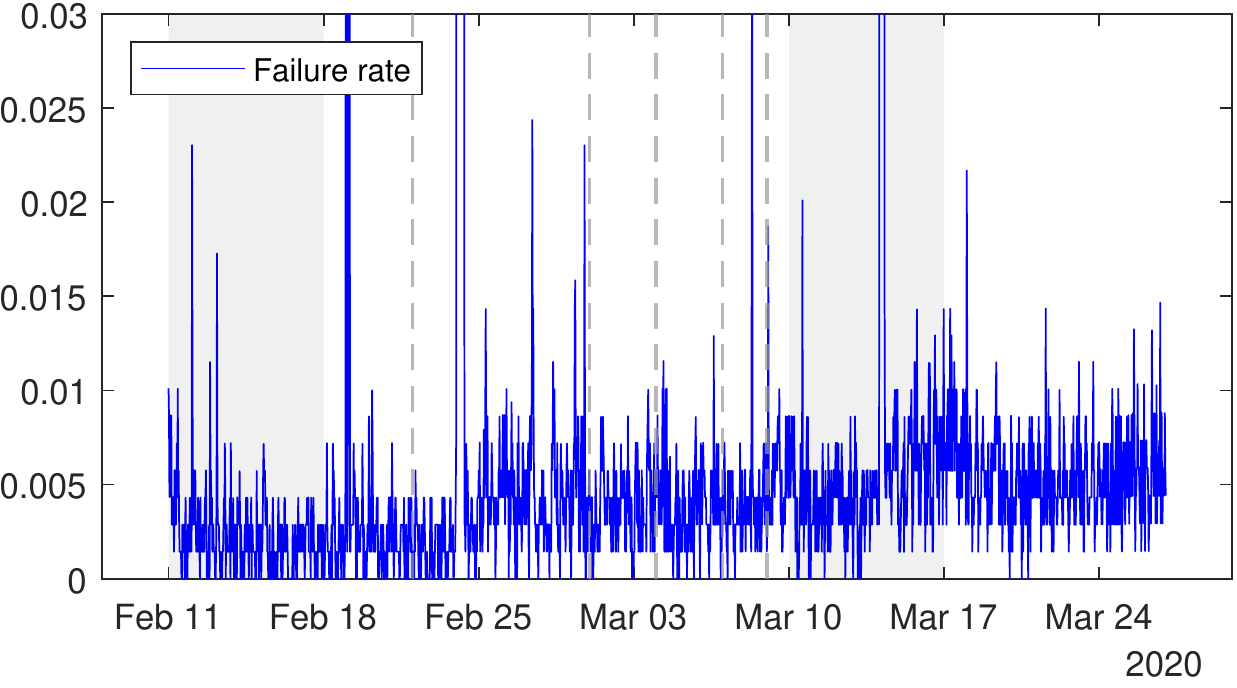}
    }
    \caption{HTTP measurements with both source and target in Italy.}
    \label{fig:http}
\end{figure*}

In the previous sections, we reported results obtained via ping measurements based on ICMP echo requests. This type of traffic is used for network diagnostics and sometimes routers can treat it differently from traffic generated by end users, even if they usually behave similarly in terms of RTT~\cite{wenwei2007evaluating}. 

Even if the trends highlighted in the previous sections by ICMP show an evident increase of the network congestion caused by the COVID-19 lockdown, we also analyze the latencies produced by RIPE Atlas HTTP Anchoring Measurements (based on TCP).
These kinds of measurements are quite different from the ICMP measurements which we analyzed above. In particular, each HTTP measurement collects just one sample (instead of at least three for the ICMP ping ones). In addition, the latencies that can be extracted from these measurements include the time for sending an HTTP request to a server and waiting for the HTTP response body to arrive. This time includes the time needed for the server to build the response. The time needed for the DNS resolution of the server address, instead, is not included. For this reason, it must be noted that the latencies extracted from HTTP measurements are not directly comparable with the latencies obtained by means of ICMP ping.

We analyzed the latencies extracted from HTTP measurements with the methodology described in Section~\ref{sec:method}, however, being each measurement composed by just one sample, we will not have $d_{min}$, $d_{avg}$, and $d_{max}$, but just $d$, and therefore just $q$ and $r$. Figure~\ref{subfig:httprtt} shows the results for measurements with source and target located in Italy. The figure shows an increase in the latency due to the lockdown. Like in the previous results, the increase is progressive and not step-like, as we recall that Italy experienced multiple partial lockdowns before entering in the strictest one. A step-like increase is noticeable in correspondence of the increase of the packet loss showed in Section~\ref{subsec:packetloss}. In this case the packet loss affects the performance of HTTP as, differently from ICMP, the packet loss triggers retransmissions of the TCP protocol. Consequently, the end-to-end latency of the TCP connection increases. Besides this anomaly, the overall increment is noticeable also by comparing average and standard deviation values of $r$, which are respectively 24.9\% and 98.5\% higher in W2 with respect to W1. In Figure~\ref{subfig:httpploss}, the failure rate of HTTP measurements is shown. Such failures are due to a combination of network errors which include connection timeout, host unreachable and network unreachable problems. Also in this case an increment of the failure rate can be noticed as long as the Italian lockdown becomes more restrictive. These results allow us to confirm that the increased latency, in Italy, during lockdown is not limited to network diagnostic traffic but also to end user traffic.

\section{Impact of COVID-19 pandemic in other European countries}
\label{sec:europe}

\begin{figure*}[t]
    \centering
    \subfloat[][AMD, from ES to ES.]{
        \includegraphics[width=0.4\textwidth]{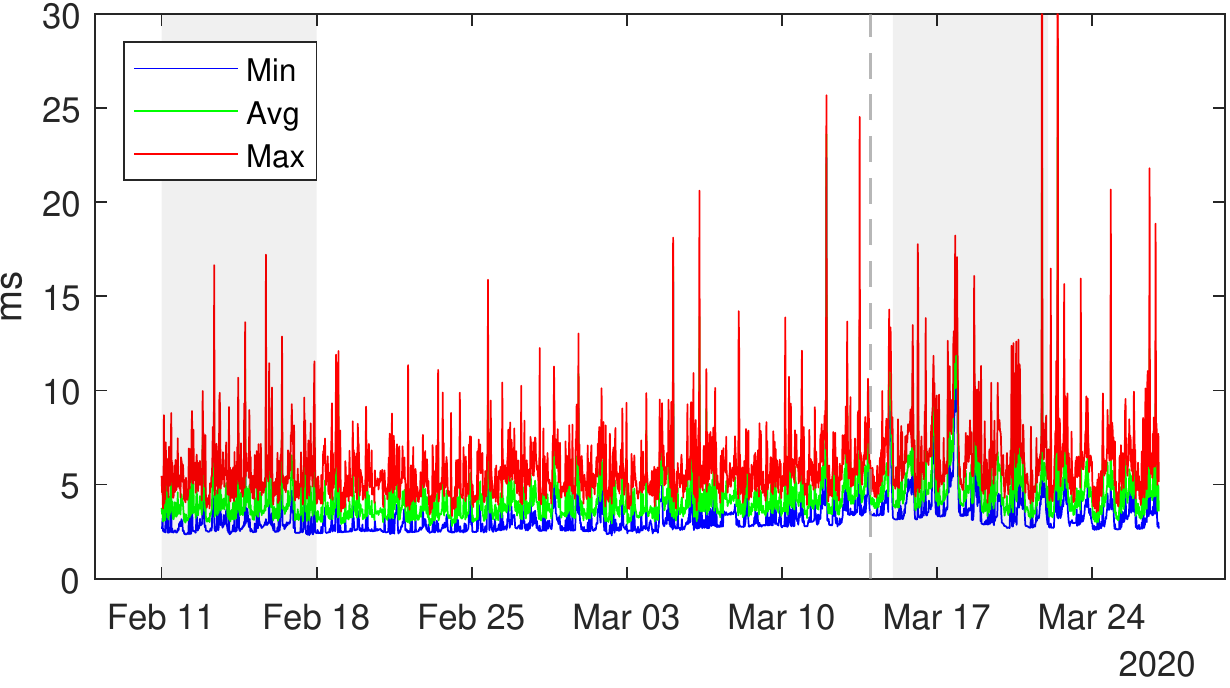}
        \label{subfig:es-es-pp-anchoring-both-all}
    }
    \subfloat[][UDMD, from ES to ES.]{
        \includegraphics[width=0.4\textwidth]{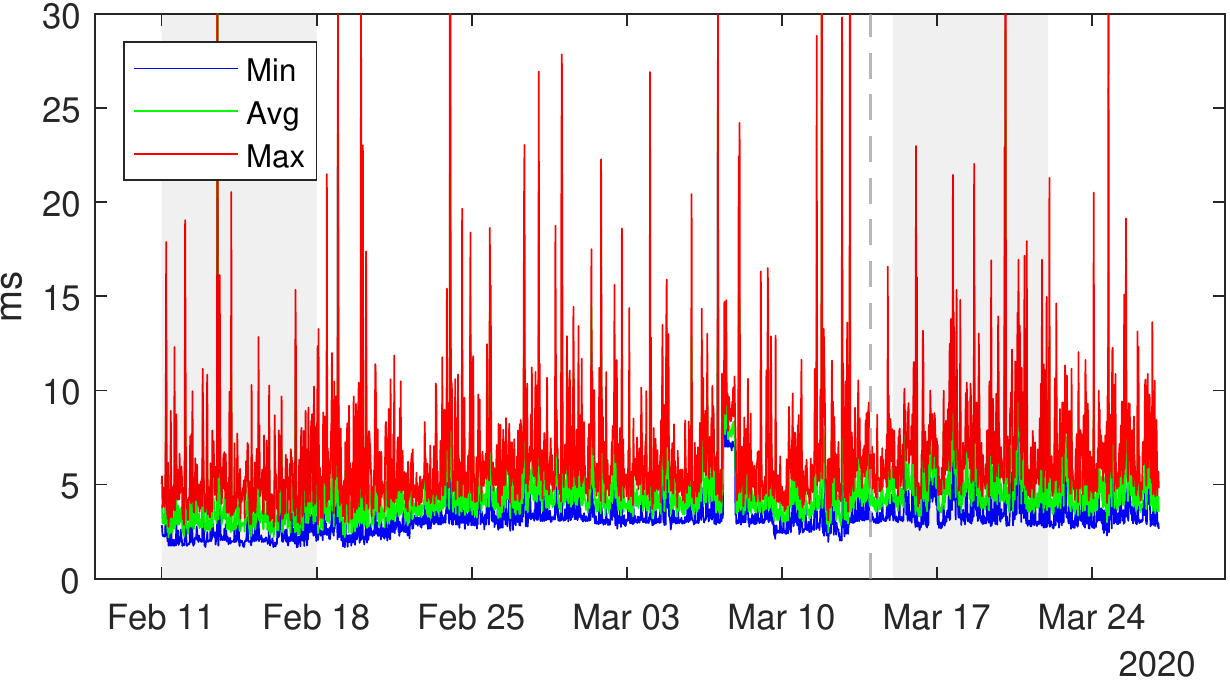}
        \label{subfig:es-es-pp-udm-both-all}
    }\\
    \subfloat[][AMD, from FR to FR.]{
        \includegraphics[width=0.4\textwidth]{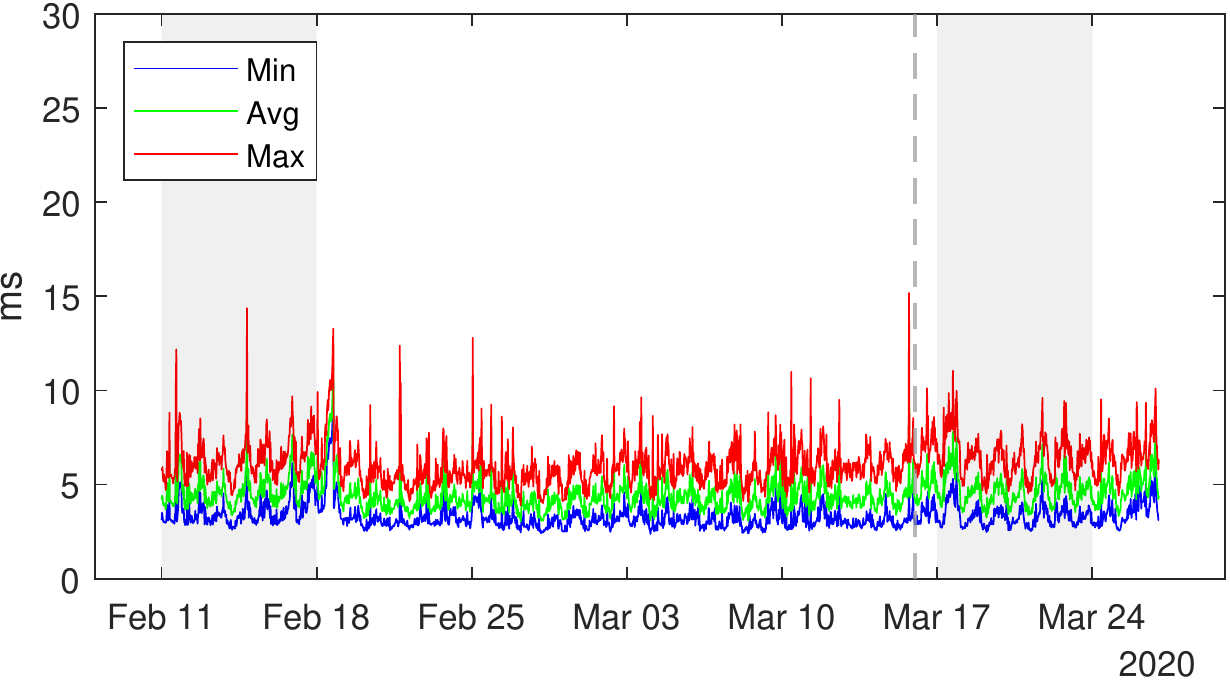}
        \label{subfig:fr-fr-pp-anchoring-both-all}
    }
    \subfloat[][UDMD, from FR to FR.]{
        \includegraphics[width=0.4\textwidth]{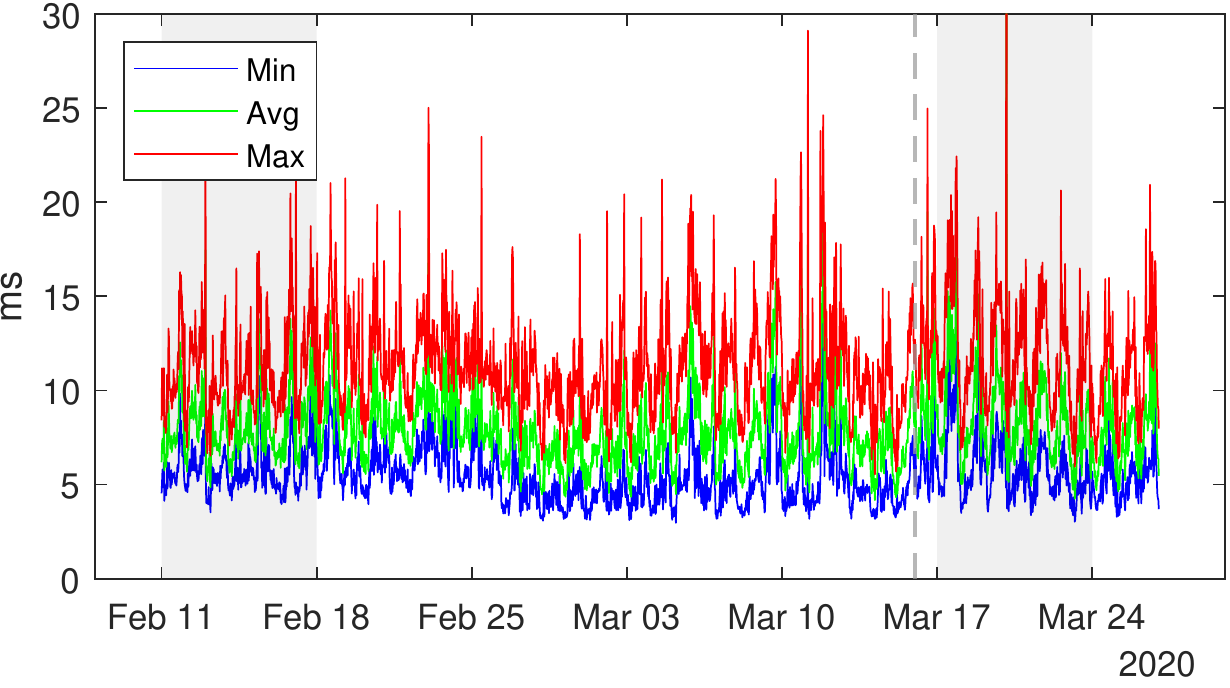}
        \label{subfig:fr-fr-pp-udm-both-all}
    }\\
    \subfloat[][AMD, from DE to DE.]{
        \includegraphics[width=0.4\textwidth]{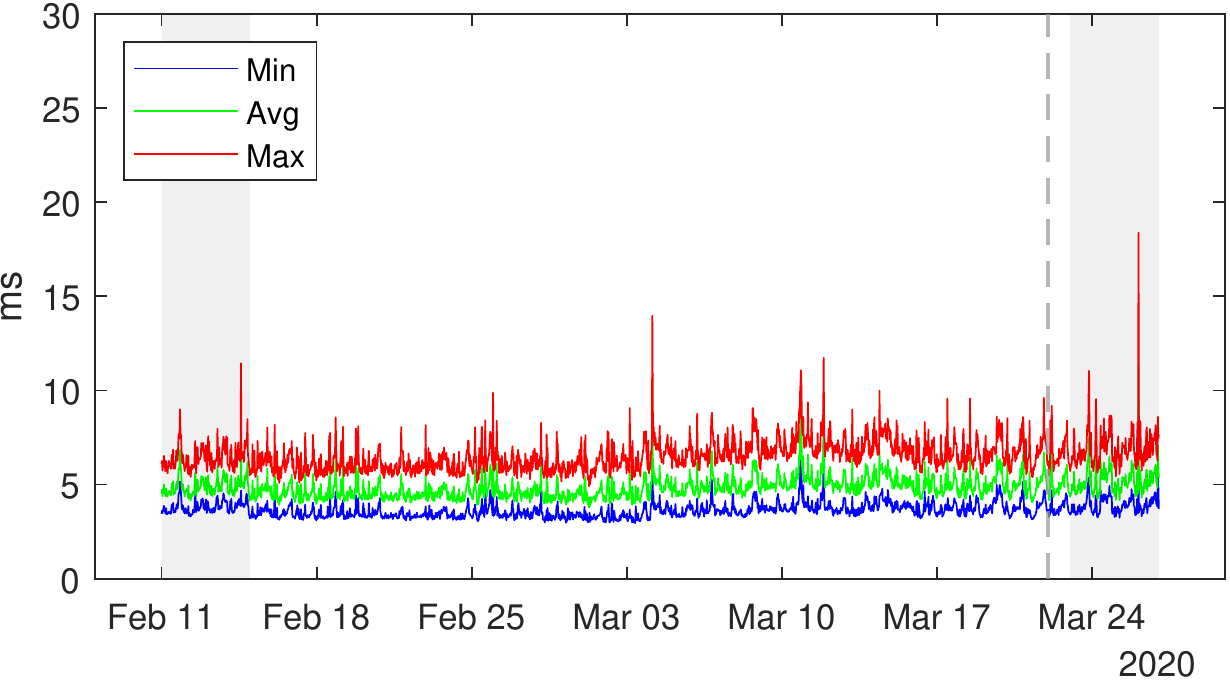}
        \label{subfig:de-de-pp-anchoring-both-all}
    }
    \subfloat[][UDMD, from DE to DE.]{
        \includegraphics[width=0.4\textwidth]{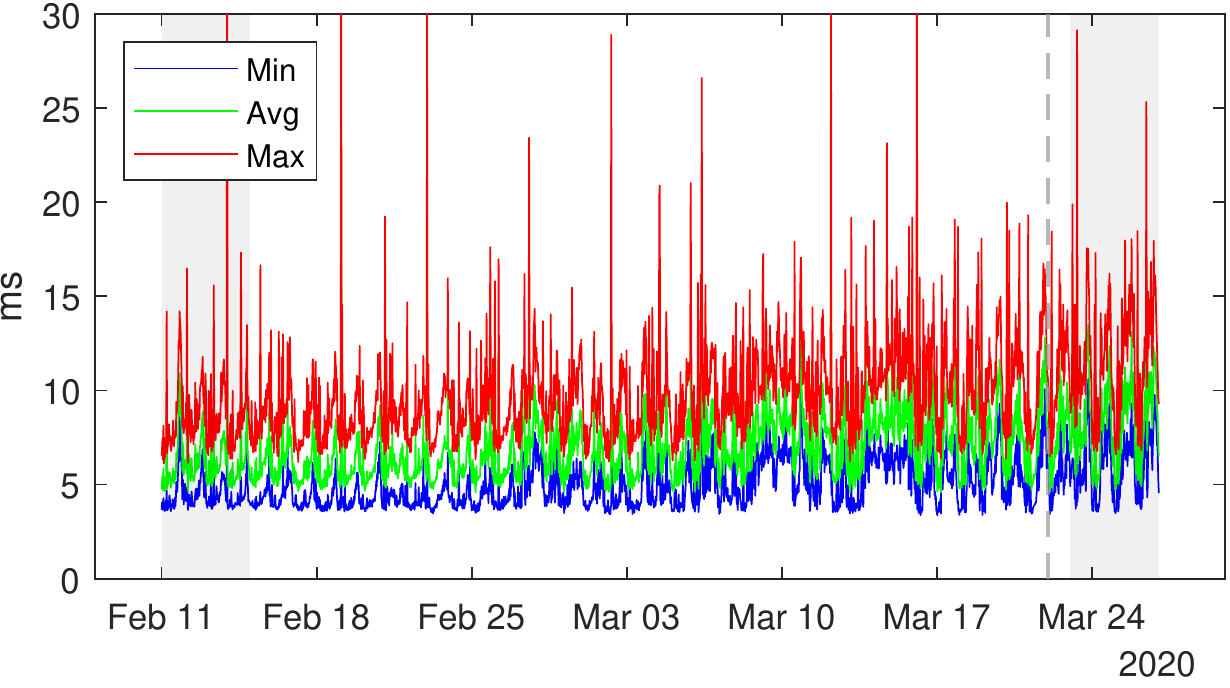}
        \label{subfig:de-de-pp-udm-both-all}
    }\\
    \subfloat[][AMD, from SE to SE.]{
        \includegraphics[width=0.4\textwidth]{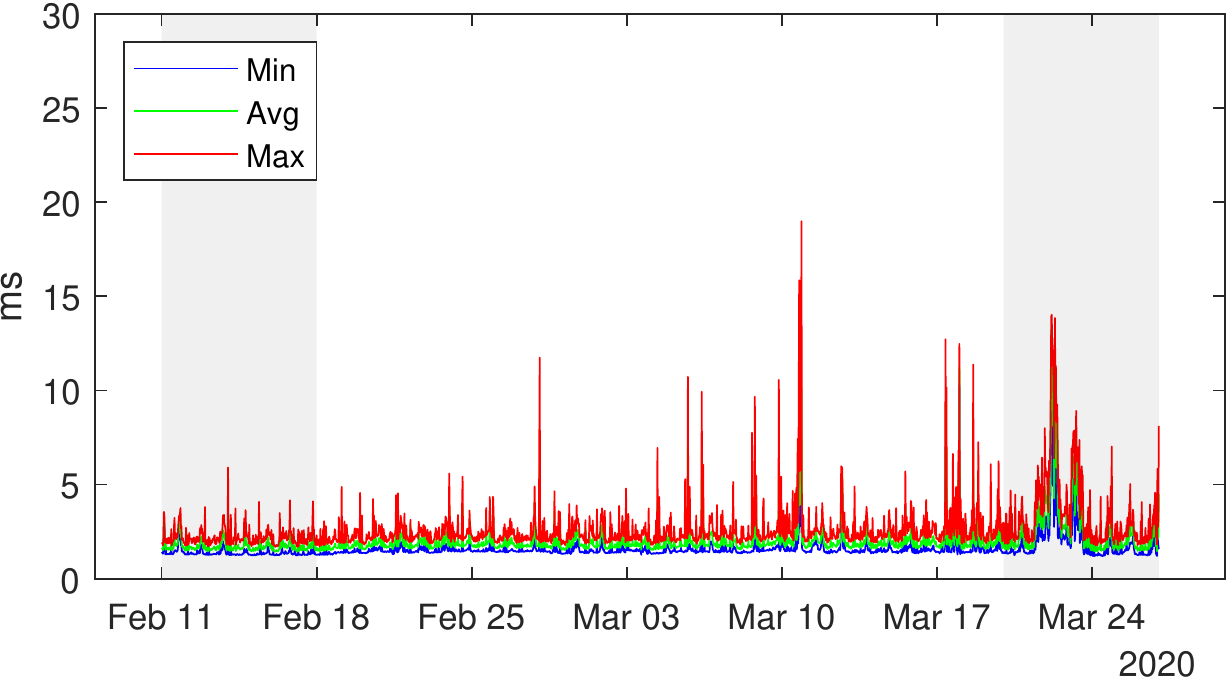}
        \label{subfig:se-se-pp-anchoring-both-all}
    }
    \subfloat[][UDMD, from SE to SE.]{
        \includegraphics[width=0.4\textwidth]{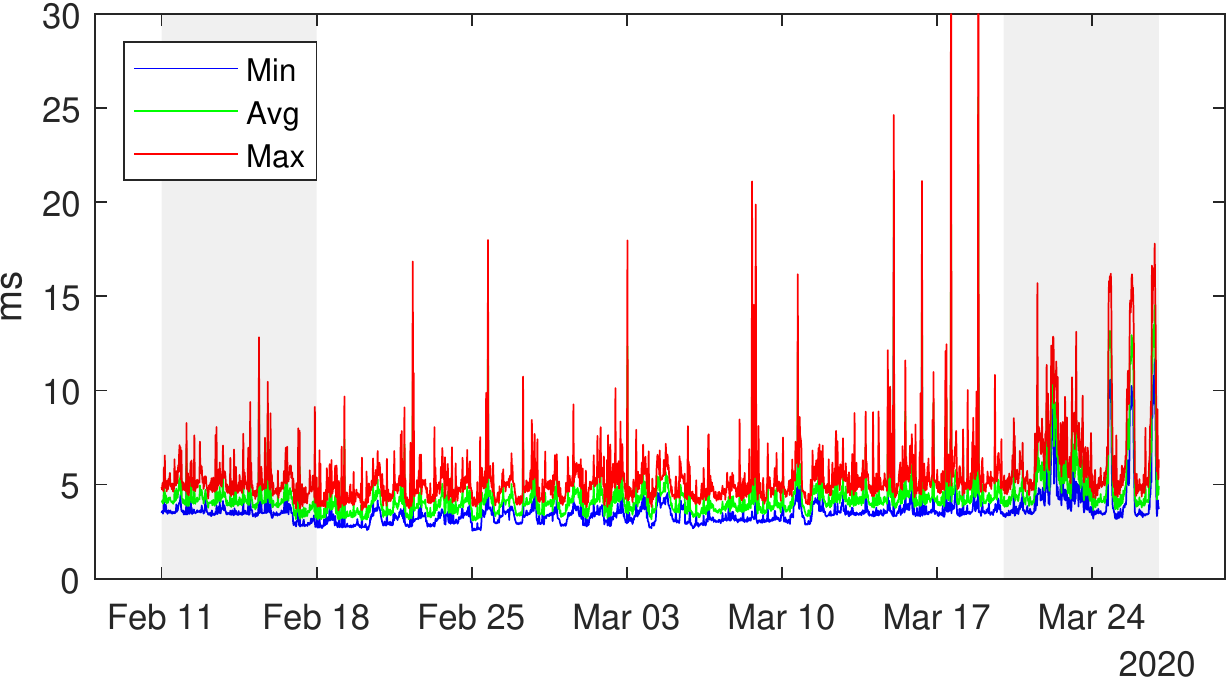}
        \label{subfig:se-se-pp-udm-both-all}
    }\\
    \subfloat[][AMD, from EU to EU.]{
        \includegraphics[width=0.4\textwidth]{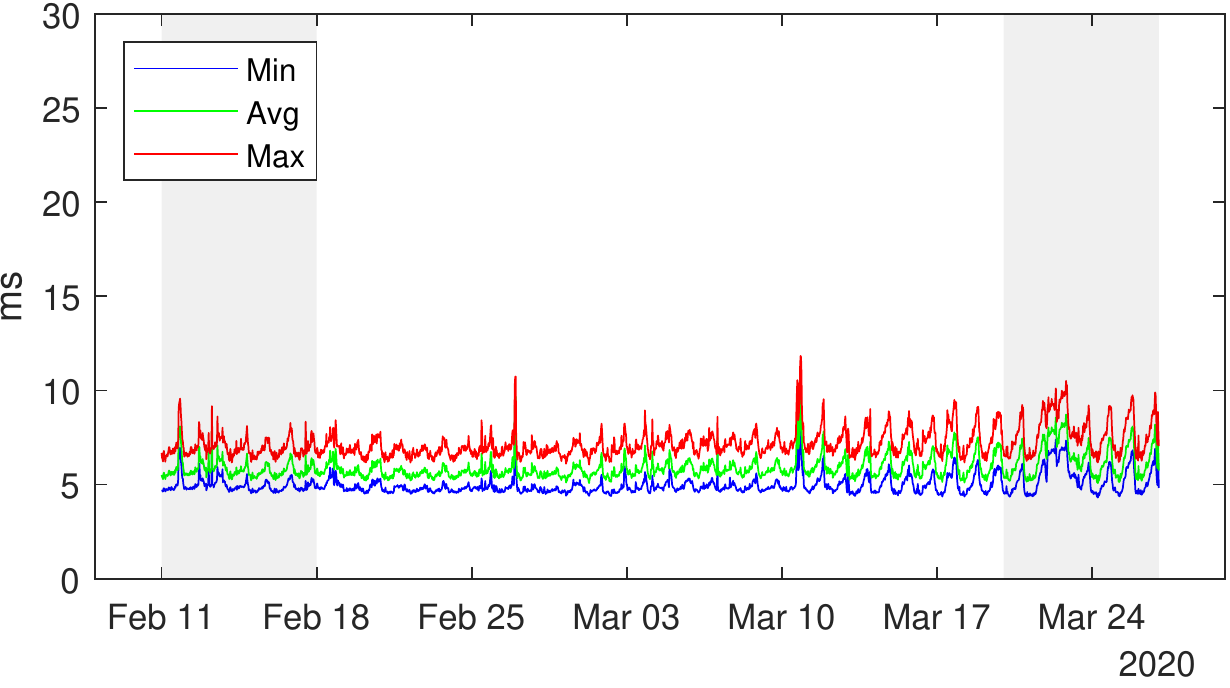}
        \label{subfig:all-all-pp-anchoring-both-all}
    }
    \subfloat[][UDMD, from EU to EU.]{
        \includegraphics[width=0.4\textwidth]{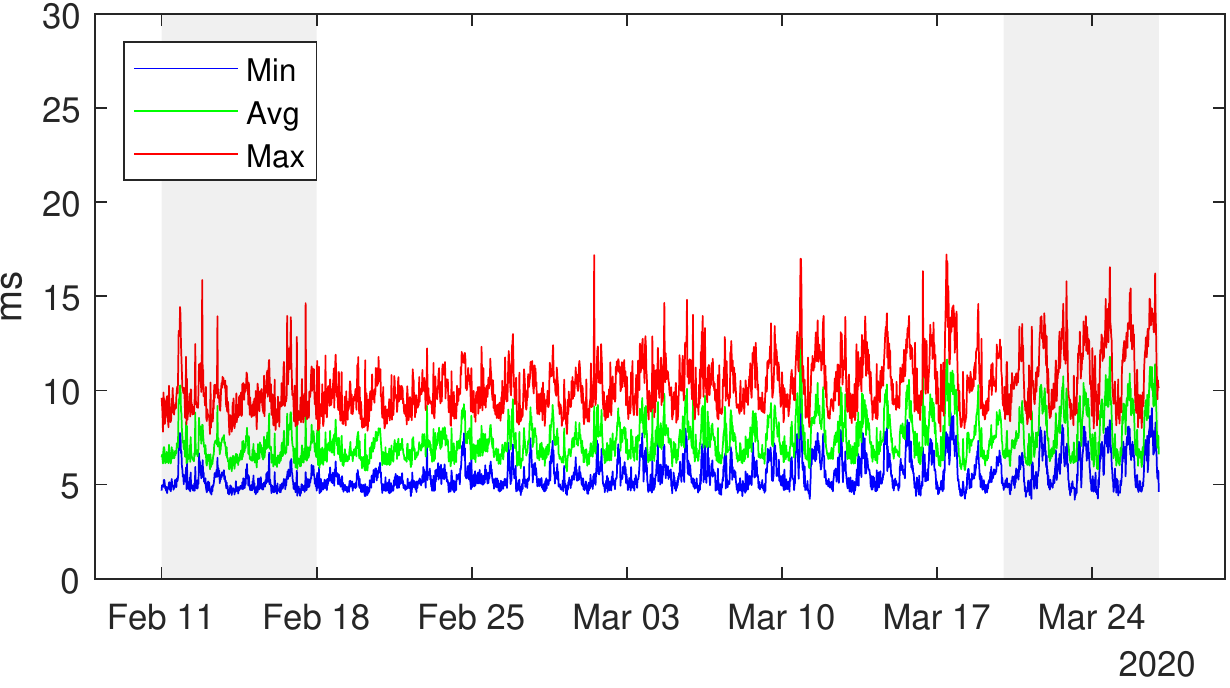}
        \label{subfig:all-all-pp-udm-both-all}
    }
    \caption{$r$ in Spain, France, Germany, Sweden and whole of Europe. The dashed vertical lines correspond to lockdown events for the considered region. The gray areas correspond to W1 and W2 for the considered region.}
    \label{fig:europe}
\end{figure*}

In this section, we show how the latency in other European countries was affected by the COVID-19 pandemic. We considered Spain, France, and Germany, which had their major lockdown events respectively five, seven, and twelve days after Italy. In addition, we considered Sweden, as an example of a country that adopted a less formal lockdown policy. Finally, we considered all European countries together. Overall results are depicted in Figure~\ref{fig:europe}.
Table~\ref{tab:europeavgstd} reports the increment of the average and standard deviation of $r_{min}$ and packet loss after stay-at-home orders. To obtain the increment, for each country we compared the first week of measurements (W1) and the first week of lockdown (W2). For Germany, the considered periods are four day long instead of seven: Germany was put on lockdown at the end of our observation period and a full week was not covered by the collected data. Since there is not a unique date for lockdown in the whole of Europe, and since some countries did not even enter a lockdown phase, for Sweden and Europe we compared the first and the last week of the observation period. The two weeks used for comparison are highlighted in light grey in Figure~\ref{fig:europe}. In the following we analyze the considered countries in detail.

Spain shows high variability of latencies also before the lockdown. However, an increase due to enforced restrictions is noticeable in AMD. Similarly, circadian patterns become more evident  (Figure~\ref{subfig:es-es-pp-anchoring-both-all}). This is confirmed also by the summary statistics, which show an increment of 37.4\% of the $r_{min}$ average, and of 110.8\% of the $r_{min}$ standard deviation, in W2 compared to W1. The visual analysis of UDMD shows a progressive increase of latency which starts much earlier than the lockdown, with a generally spread variability (Figure~\ref{subfig:es-es-pp-udm-both-all}). In fact, results in Table~\ref{tab:europeavgstd} show a significant increment in the average $r_{min}$ of 63.2\%, but a modest 19.3\% for the standard deviation. On March 7, a temporary increase can be observed in Figure~\ref{subfig:es-es-pp-udm-both-all}. We investigated this phenomenon and found that it is due to a considerable increase in the latency towards one target, observed from multiple sources. We believe this to be an anomalous behavior due to the target itself or the network in its proximity. The analysis of the packet loss shows opposite behaviours between AMD and UDMD. In AMD, the packet loss decreases during the lockdown, while in UDMD increases greatly.

In France, the situation is different, as shown in Figures~\ref{subfig:fr-fr-pp-anchoring-both-all} and~\ref{subfig:fr-fr-pp-udm-both-all}, which show AMD- and UDMD-based results respectively. In AMD, the overall increase of latency is barely noticeable. Lockdown seems to accentuate the periodic fluctuations due to circadian rhythms. In fact, the summary statistics included in Table~\ref{tab:europeavgstd} show even a decrement of $r_{min}$ average and standard deviation in W2 compared to W1, in the AMD. However, it must be noticed that the first week of measurements in AMD appears as particularly noisy if compared to the other pre-lockdown weeks. UDMD instead show much higher variability, which is the attribute that shows the most evident impact of the lockdown. In fact, the analysis shows an increment of just 0.5\% for the $r_{min}$ average and 49.6\% for the standard deviation of $r_{min}$. The packet loss increases during lockdown for both AMD and UDMD, in both average and standard deviation.

In Germany the situation is different from both Spain and France. In AMD, the effect of lockdown is noticeable only in terms of amplified circadian patterns (Figure~\ref{subfig:de-de-pp-anchoring-both-all}). The analysis confirms this, by showing just a 2.8\% increase of the $r_{min}$ average, and a 37.3\% increase of the $r_{min}$ standard deviation. In UDMD, a more significant increase of latency is visible (Figure~\ref{subfig:de-de-pp-udm-both-all}), which corresponds to a 34.1\% higher value for the $r_{min}$ average and a 88.7\% higher value for the $r_{min}$ standad deviation in W2 compared to W1. It is worth noticing that in Germany the degradation of latency starts more than one week before the lockdown. This happened because Germany, like Italy, proceeded to some partial lockdowns and school closures before the major restrictions. The average packet loss slightly increases for UDMD, and decreases for AMD.

As mentioned, Sweden adopted less formal restrictions. For this reason, results shown in Figures~\ref{subfig:se-se-pp-anchoring-both-all} and~\ref{subfig:se-se-pp-udm-both-all} are definitely interesting. Both AMD and UDMD show a progressive increase of $r$ and of its variability in the considered time period. This can mean either that Swedish people autonomously increased social distancing and implemented stay-at-home policies as suggested by the Swedish government, or that the performance of the Swedish Internet infrastructure has been affected by the lockdown imposed in other countries. Also the comparison of W1 and W2 show a significant increase of $r_{min}$ average and standard deviation: 44.7\% and 531.2\% for AMD and 28.6\% and 231.7\% for UDMD. The packet loss increases just for AMD, but decreases for UDMD.

Figure~\ref{subfig:all-all-pp-anchoring-both-all} and Figure~\ref{subfig:all-all-pp-udm-both-all} show respectively the AMD- and UDMD-based results for all of Europe. Additional latency is generally smaller than the individual countries we analyzed. The response to national lockdowns seems to be fairly good even if we notice an accentuation of the variability due to the circadian activities, starting from the Italian lockdown, but not a significant increase of the overall latencies. This is confirmed by the statistics reported in Table~\ref{tab:europeavgstd}: the $r_{min}$ average is subject to a modest increase, equal to 8.1\% and 15.4\% in AMD and UDMD respectively, while the $r_{min}$ standard deviation experiences a significant increase, equal to 114.6\% and 108.1\%. These results seem to indicate that, on a continent-level scale, the impact of lockdown is still noticeable but without dramatic changes in observed performance. The packet loss instead decreases for AMD and is almost unchanged for UDMD.

It is worth to notice that, after analyzing the packet loss in the different countries, we cannot conclude that an increase in latency is coupled with an increase in packet loss. In fact, in some cases we found increasing latency and decreasing packet loss, and vice-versa.

\begin{table*}[t]
    \centering
    \caption{$r_{min}$ and packet loss average and standard deviation increments (W2 compared to W1) in Italy, Spain, France, Germany, Sweden, and whole of Europe.}
    \footnotesize
    \begin{tabular}{|l|r|r|r|r|r|r|r|r|}
    \hline
    {\it Country} & \multicolumn{2}{|c|}{\it $r_{min}$ avg. increment (\%)} & \multicolumn{2}{|c|}{\it $r_{min}$ std. dev. increment (\%)} & \multicolumn{2}{|c|}{\it packet loss avg. increment (\%)} & \multicolumn{2}{|c|}{\it packet loss std. dev. increment (\%)} \\
    \hline
    & {\it AMD} & {\it UDMD} & {\it AMD} & {\it UDMD} & {\it AMD} & {\it UDMD} & {\it AMD} & {\it UDMD}\\
    \hline
    Italy & 27.7 & 66.8 & 203.6 & 280.8 & 110.3 & 205.7 & 507.2 & 367.9 \\
    \hline
    Spain & 37.4 & 63.2 & 110.8 & 19.3 & -55.6 & 134.3 & -36.6 & 49.4 \\
    \hline
    France & -5.8 & 0.5 & -15.6 & 49.6 & 17.0 & 43.5 & 45.4 & 37.4 \\
    \hline
    Germany & 2.8 & 34.1 & 37.3 & 88.7 & -22.7 & 18.3 & -33.8 & -14.6 \\
    \hline
    Sweden & 44.7 & 28.6 & 531.2 & 231.7 & 101.0 & -33.7 & 52.8 & -47.6 \\
    \hline
    Europe & 8.1 & 15.4 & 114.6 & 108.1 & -21.6 & -0.3 & -18.5 & 0.6 \\
    \hline
    \end{tabular}
    \label{tab:europeavgstd}
\end{table*}

\section{Conclusion}
\label{sec:conclusion}

It is well-known that computer viruses may cause a slowdown of the Internet \cite{10.1145/66093.66095,1306386}. In 2020 we all learned that also biological viruses may affect the global Internet performance, because of the changes they bring in the way we live. 

In this paper, we analyzed the impact of the COVID-19 pandemic on the latency of the Internet on a large scale. Latency is particularly important not only because it has a profound effect on some classes of applications, but also because it is, by itself, an excellent indicator of the health status of the network. Results, which have been obtained from the analysis of a large amount of measurements, show that the impact of the increased on-line activities is relevant, especially in terms of higher variability. The major changes have been observed in the evening, the time of the day when most of the on-line activities are related to entertainment. This suggests that distance learning and remote working contributed to a lesser extent in terms of additional network latency. Results obtained for the considered countries show relevant differences, which can be due to the resilience levels of their network and/or to the non-uniform restrictions imposed by authorities.

We believe that the provided numbers and the related analysis, despite being limited to a portion of the Internet, definitely help in better understanding this previously unseen event in the history of the Internet.

\bibliographystyle{elsarticle-num}
\bibliography{main}

\end{document}